\documentclass[aps,prc,twocolumn,nofootinbib,superscriptaddress,groupedaddress]{revtex4}
\usepackage{dcolumn}   
\usepackage{bm}        
\usepackage{multirow}
\usepackage{xcolor}

\usepackage{amsmath}
\usepackage{amsfonts}
\usepackage{amssymb}
\usepackage{makeidx}
\usepackage{graphicx}
\usepackage{anysize}
\usepackage{hyperref}
\usepackage{slashed}

\newenvironment{mymathbox}
{\par\smallskip\centering\begin{lrbox}{0}%
\begin{minipage}[c]{0.8\textwidth}}
{\end{minipage}\end{lrbox}%
\framebox[0.9\textwidth]{\usebox{0}}%
\par\medskip
\ignorespacesafterend}
\newcommand{\bb}{\begin{mymathbox}}
\newcommand{\eb}{\end{mymathbox}}

\newcommand{\be}{\begin{equation}}
\newcommand{\ee}{\end{equation}}
\newcommand{\ba}{\begin{eqnarray}}
\newcommand{\ea}{\end{eqnarray}}

\newcommand{\npsi}{{\bf \npsi}}

\newcommand{\bma}{\begin{pmatrix}}
\newcommand{\ema}{\end{pmatrix}}

\begin{document}

\title{Constraints in modeling the quasielastic response in inclusive lepton-nucleus scattering}

\author{R.~Gonz\'alez-Jim\'enez}
\affiliation{Grupo de F\'isica Nuclear, Departamento de Estructura de la Materia, F\'isica T\'ermica y Electr\'onica, Facultad de Ciencias F\'isicas, Universidad Complutense de Madrid and IPARCOS, CEI Moncloa, Madrid 28040, Spain\looseness=-1}
\author{M.B.~Barbaro}
\affiliation{Dipartimento di Fisica, Universit\`{a} di Torino and INFN, Sezione di Torino, Via P. Giuria 1, 10125 Torino, Italy\looseness=-1}
\author{J.A.~Caballero}
\affiliation{Departamento de F\'{i}sica At\'omica, Molecular y Nuclear, Universidad de Sevilla, 41080 Sevilla, Spain\looseness=-1}
\author{T.W.~Donnelly}
\affiliation{Center for Theoretical Physics, Laboratory for Nuclear Science and Department of Physics, Massachusetts Institute of Technology, Cambridge, Massachusetts 02139, USA\looseness=-1}
\author{N.~Jachowicz}
\affiliation{Department of Physics and Astronomy, Ghent University, Proeftuinstraat 86, B-9000 Gent, Belgium\looseness=-1}
\author{G.D.~Megias}
\affiliation{Departamento de F\'{i}sica At\'omica, Molecular y Nuclear, Universidad de Sevilla, 41080 Sevilla, Spain\looseness=-1}
\affiliation{IN2P3-CNRS, Laboratoire Leprince-Ringuet, Palaiseau 91120, France\looseness=-1}
%
\author{K.~Niewczas}
\affiliation{Institute of Theoretical Physics, University of Wroc{\l}aw, Plac Maxa Borna 9, 50-204, Wroc{\l}aw, Poland\looseness=-1}
\affiliation{Department of Physics and Astronomy, Ghent University, Proeftuinstraat 86, B-9000 Gent, Belgium\looseness=-1}
\author{A.~Nikolakopoulos}
\affiliation{Department of Physics and Astronomy, Ghent University, Proeftuinstraat 86, B-9000 Gent, Belgium\looseness=-1}
\author{J.M.~Ud\'ias}
\affiliation{Grupo de F\'isica Nuclear, Departamento de Estructura de la Materia, F\'isica T\'ermica y Electr\'onica, Facultad de Ciencias F\'isicas, Universidad Complutense de Madrid and IPARCOS, CEI Moncloa, Madrid 28040, Spain\looseness=-1}

\date{\today}

\begin{abstract}
We show that the quasielastic (QE) response calculated with the SuSAv2 (superscaling approach) model, that relies on the scaling phenomenon observed in the analysis of $(e,e')$ data and on the relativistic mean-field theory, is very similar to that from a relativistic distorted wave impulse approximation model when only the real part of the optical potentials is employed. 
The coincidence between the results from these two completely independent approaches, which satisfactorily agree with the inclusive data, reinforces the reliability of the quasielastic predictions stemming from both models and sets constraints for the QE response. 
We also study the low energy and momentum transfer region of the inclusive response by confronting the results of the relativistic mean-field model with those of the Hartree-Fock continuum random-phase approximation model, which accounts for nuclear long-range correlations. 
Finally, we present a comparison of our results with the recent JLab $(e,e')$ data for argon, titanium and carbon, finding good agreement with the three data sets.
\end{abstract}

\maketitle

\section{Introduction}

The study of electron-nucleus scattering continues to be the main source of knowledge for the understanding of neutrino-nucleus interactions. The vector part of the interaction can be inferred directly from electron scattering and the influence of the nuclear medium is identical in both processes. Consequently, the comparison with electron scattering data 
is a necessary test for any theoretical approach aiming at modeling the neutrino-nucleus interaction.

Many different reaction channels contribute to the lepton-nucleus cross section. Looking at the primary vertex, the lepton can interact, depending on the transferred momentum and energy, via elastic scattering, collective nuclear excitations, discrete resonances, quasielastic (QE) scattering, multinucleon knockout processes, one- and two-pion production, and other processes typically encompassed in the deep-inelastic scattering (DIS) response. On top of that, the secondary interactions of the outgoing hadron(s) with the residual system may lead to very complex final states with high hadron multiplicities.
The final goal of Monte Carlo (MC) neutrino event generators (NuWro~\cite{Golan12}, GENIE~\cite{Andreopoulos10}, NEUT~\cite{Hayato09}, and GiBUU~\cite{Buss12}) is to model all these reactions.
A simpler case is the inclusive process, defined as the one in which only the scattered lepton is detected.
In neutrino-oscillation experiments, this is the main signal in the far Cherenkov-like detectors~\cite{HyperK15}. 
Hence, a minimum requirement for MC generators is that, after integration over the hadron (undetected) variables, they should be able to provide ``good'' inclusive results.

Many investigations have been devoted to the study of the inclusive response for several decades~\cite{Foundations17,Alvarez-Ruso18}. 
This is consistent with the fact that, from a theoretical point of view, the description of inclusive processes is simpler than other scenarios where specific knowledge of the final hadron multiplicities is needed, and consequently one needs to account for the effects produced by the propagation of the knock-out nucleon through the nuclear medium. 
On the contrary, the inclusive case can be properly described by simply computing the self-energy of the propagating nucleon without the need to track the details of the secondary excitations in the final state. 
This can be done efficiently by using the mean-field based models with the final nucleon described as a wave function distorted by the average potential produced by the residual nucleus. 
Although this formalism is entirely based on the impulse approximation (IA), namely, one-body currents and single-particle equations, the model incorporates effective potentials that can account for effects beyond the mean field captured from the analysis of data. This is the case of the phenomenological energy-dependent complex optical potentials fitted to nucleon-nucleus scattering data. The inclusive responses, {\it i.e.,} no flux lost, can be handled by simply removing the imaginary (absortive) terms in the potentials. For the quasielastic peak, this yields results that are numerically very similar to the, formally more sound, incorporation of inelasticities represented in the full complex optical potential which is acomplished by the Relativistic Green Function (RGF)~\cite{Capuzzi91,Meucci09,Ivanov16b}. Here the flux lost into inelastic channels, represented by the imaginary term of the optical potential fitted to proton-nucleus elastic scattering, is recovered by a formal summation on those inelastic channels. Other descriptions of the inclusive responses are based on the use of spectral function models plus convolution approach~\cite{Ankowski15a,Ivanov19}, the local Fermi gas approach including random phase approximation (RPA) correlations~\cite{Nieves11,Martini10,Martini11}, the Green function Monte Carlo method~\cite{Rocco17}, and the scaling properties fulfilled by $(e,e')$ data~\cite{Amaro05a,Gonzalez-Jimenez14b,Ivanov16a,Amaro18}. 
 
It is important to point out that the extremely complicated many-body coupled-channel configurations contributing to the inclusive scattering cross sections make it extremely difficult to solve the problem consistently. One needs to resort to approximate models that emphasize different degrees of freedom and account for several ingredients of the process. Although these approaches build the inclusive responses out of many different contributions, they can lead to similar results. In most of the cases, the various contributions to the inclusive signal can significantly overlap with each other making it difficult to experimentally separate the different reaction channels. This is the case, for instance, for the QE and the two-nucleon knockout responses. This difficulty to disentangle without ambiguity the role of each contribution explains why different models, with very different ingredients, can produce similar inclusive results.
As an example, the predictions from GiBUU in~\cite{Mosel19} and those from the SuSAv2 model of \cite{Barbaro19} agree to a large extent when compared to inclusive $(e,e')$ data, while each contribution separately, e.g. the QE and 2 particle-2 hole channels, show important discrepancies.

In this work we focus on QE scattering. We show the results of the SuSAv2 and different mean-field based models at low, intermediate and high values of the momentum transfer $q$. We aim at pointing out under which conditions they fail or they do a good job, explaining the reasons in each case. Finally, by comparing the results of these different approaches over a broad energy range, we are able to set constraints on the modeled QE responses.

The outline of this work is as follows. In Sect.~\ref{Models} we illustrate the models. In Sect.~\ref{Scaling-functions} we show and discuss the scaling functions obtained with the different approaches. In Sect.~\ref{JLab-comparison} we compare our predictions with recent $(e,e')$ JLab data. Our conclusions are presented in Sec.~\ref{Conclusions}.

\section{Models}\label{Models}

All of the models employed here are based on the impulse approximation, namely the primary interaction is assumed to be that of the probe with a single bound nucleon that is knocked out and further propagates in the nuclear medium, experiencing final-state interactions (FSI).
In what follows we will differentiate between mean-field models and SuSAv2. 

In the mean-field models, the inclusive differential cross section is obtained from the exclusive one by integration over the final nucleon variables and performing a sum over the occupied shells. 
In the relativistic models the fixed-energy Dirac equation is solved using scalar and vector potentials, while in the Hartree-Fock (HF) continuum random-phase approximation (CRPA) approach, the single particle Schr\"odinger equation with the Skyrme-based HF potential is used. 

The {\it relativistic mean-field models} presented in this work differ only in the treatment of the knockout nucleon wave function. These are summarized in what follows: 
\begin{itemize}
 \item RPWIA (relativistic plane-wave impulse approximation): The final nucleon is described by a relativistic plane wave.
 \item RPWIA$(p_N>k_F)$: The final nucleon is a relativistic plane wave but if the nucleon has a momentum lower than a given Fermi momentum $k_F$, the cross section is set to zero. 
This is the way of accounting for the Pauli blocking in Fermi gas based models, where the nucleons are labeled by their momenta. In models for finite nuclear matter, however, this procedure is not consistent because nucleons have well-defined energy and angular momentum quantum numbers but they are not momentum eigenstates, so Pauli blocking should be ensured by simply using orthogonal states.
More details can be found in \cite{Gonzalez-Jimenez19}.
 \item PB-RPWIA (Pauli-blocked RPWIA): The final nucleon wave function is constructed as a plane wave that has been corrected for the overlap with the initial state. 
Thus, initial and final states are orthogonal and this removes spurious contributions to the inclusive result. 
More details can be found in \cite{Gonzalez-Jimenez19}.
 \item RMF (relativistic mean-field): The final nucleon is a scattering solution of the same Dirac equation used to describe the bound state. 
 Hence, orthogonality and rescattering of the nucleon in the final state are naturally included. 
 This potential is energy-independent which, eventually, results in modifications of the QE response  that are too large at high energies, when compared with the data. The phenomenology tells us that the interaction of the nucleon with the residual nuclear system should weaken at large energies; this would force the RMF potential to include some energy dependence, as will be done in the ED-RMF model described below.
 \item EDAD1 and EDAI: The final nucleon travels under the effect of phenomenological relativistic optical potentials that were adjusted to reproduce elastic proton-nucleus scattering data in the range $22<T_p<1100$ MeV, with $T_p$ the kinetic energy of the proton in the lab frame~\cite{Cooper93,Cooper09}.  
 The energy-dependent A-dependent fit-1 potential (EDAD1) was fitted to reproduce scattering data on a large set of target nuclei, from carbon to lead.
 The energy-dependent A-independent potentials were fitted to proton-carbon (EDAI-C) and proton-calcium (EDAI-Ca) data only.
 In this work, we focus on the inclusive $(e,e')$ cross section, where the contributions from all inelastic channels should be retained: therefore, to be consistent with flux conservation, we take only the real part of these potentials~\cite{Maieron03,Kim03,Caballero05,Meucci09,Kim07,Butkevich07}.
 Notice that the strength of the potential gets smaller at larger energies. 
 \item ED-RMF (energy-dependent RMF): The potentials describing the motion of the knocked-out nucleon are the RMF scalar (S) and vector (V) potentials but multiplied by a ``blending function'' $f_b(T_N)$ that scales them down  as the kinetic energy of the scattered nucleon increases (Fig.~\ref{fig:function}). The function $f_b(T_N)$ is inspired by the SuSAv2 analysis presented in \cite{Megias16a} and is explained in \cite{Gonzalez-Jimenez19}. 
 This function is parametrized as follows:
 $f_b(T_N) = L(T_N) + F(T_N)$, with $L$ and $F$ a Lorenztian- and Fermi-like functions given by 
 \ba
   L(T_N) =& \frac{0.85}{(T_N/200)^2 + 3.5} + 0.29\,,\\
   F(T_N) =& \frac{0.48}{ \exp[(T_N-90)/23] + 1}\,.
 \ea 
 $T_N$ is in the lab frame and in MeV. 

\begin{figure}[htbp]
  \centering
      \includegraphics[width=.26\textwidth,angle=270]{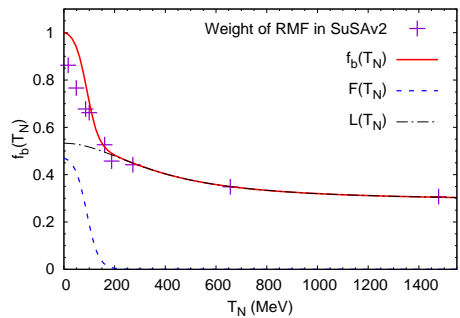}
  \caption{Blending function in the ED-RMF model. The crosses correspond to the weight of the RMF contribution in the SuSAv2~\cite{Megias16a}, as explained in~\cite{Gonzalez-Jimenez19}. }
  \label{fig:function}  
\end{figure}
  
 This approach is similar to the EDAD1 and EDAI discussed above in the sense that it uses real potentials that decrease with energy, however, it presents some advantages:
 i) for knockout nucleons with small energies ($T_p<100$ MeV), consequently when the overlap between initial and final state is non-negligible, the ED-RMF potentials essentially coincide with the original RMF ones, preventing non-orthogonality issues (similar approaches were employed in~\cite{Kim07,Ivanov16b});
 ii) for scattered nucleons with larger energies ($T_p>1100$ MeV) the potentials tend to a minimum value (following the behavior suggested by the SuSAv2 model). 
 This avoids the problem of the optical potentials when they need to be used outside the region of the fit. 
 The ED-RMF is explained in more detail in \cite{Gonzalez-Jimenez19}. 
\end{itemize}

\begin{figure*}[htbp]
  \centering
      \includegraphics[width=.23\textwidth,angle=270]{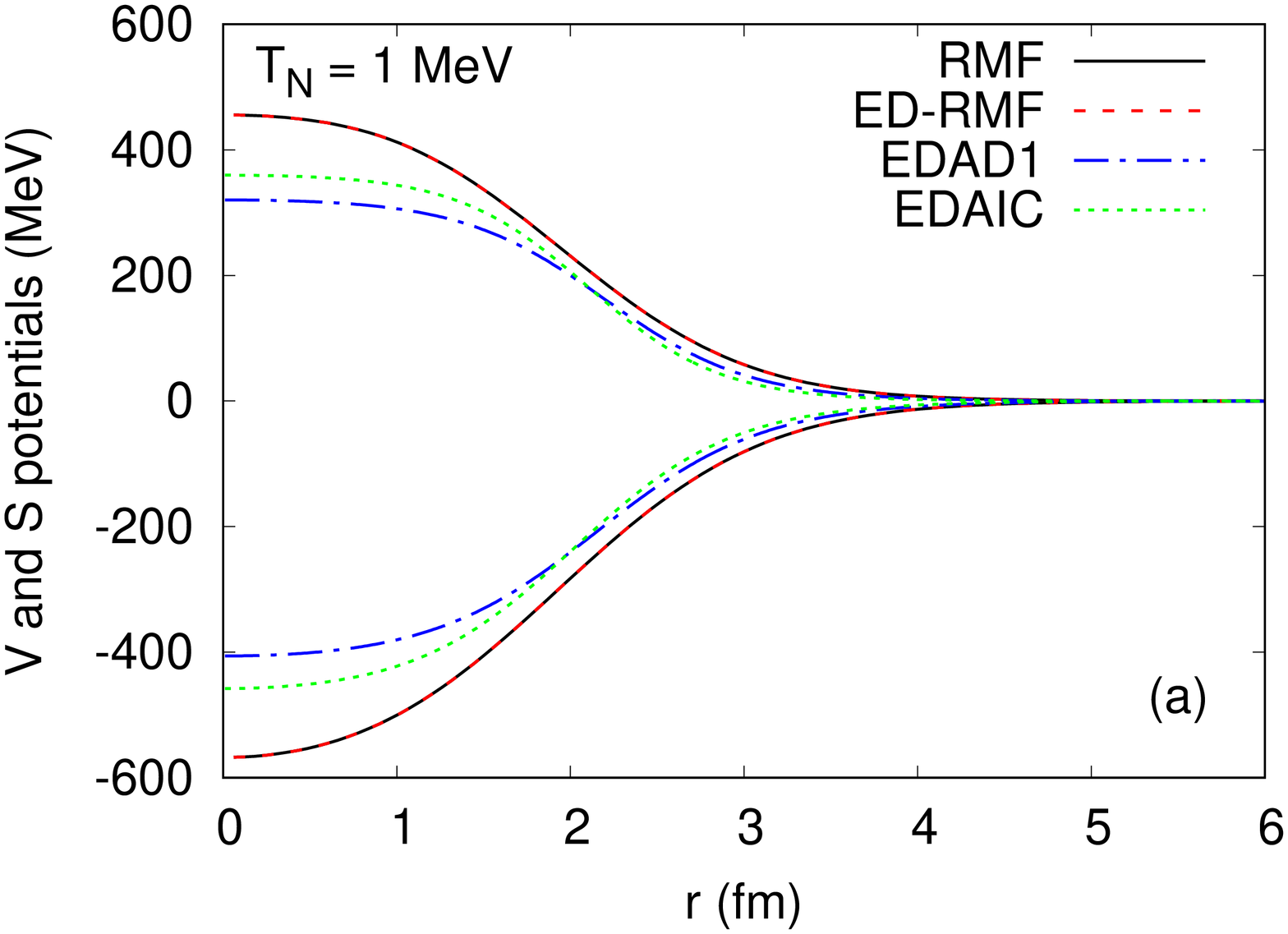}
      \includegraphics[width=.23\textwidth,angle=270]{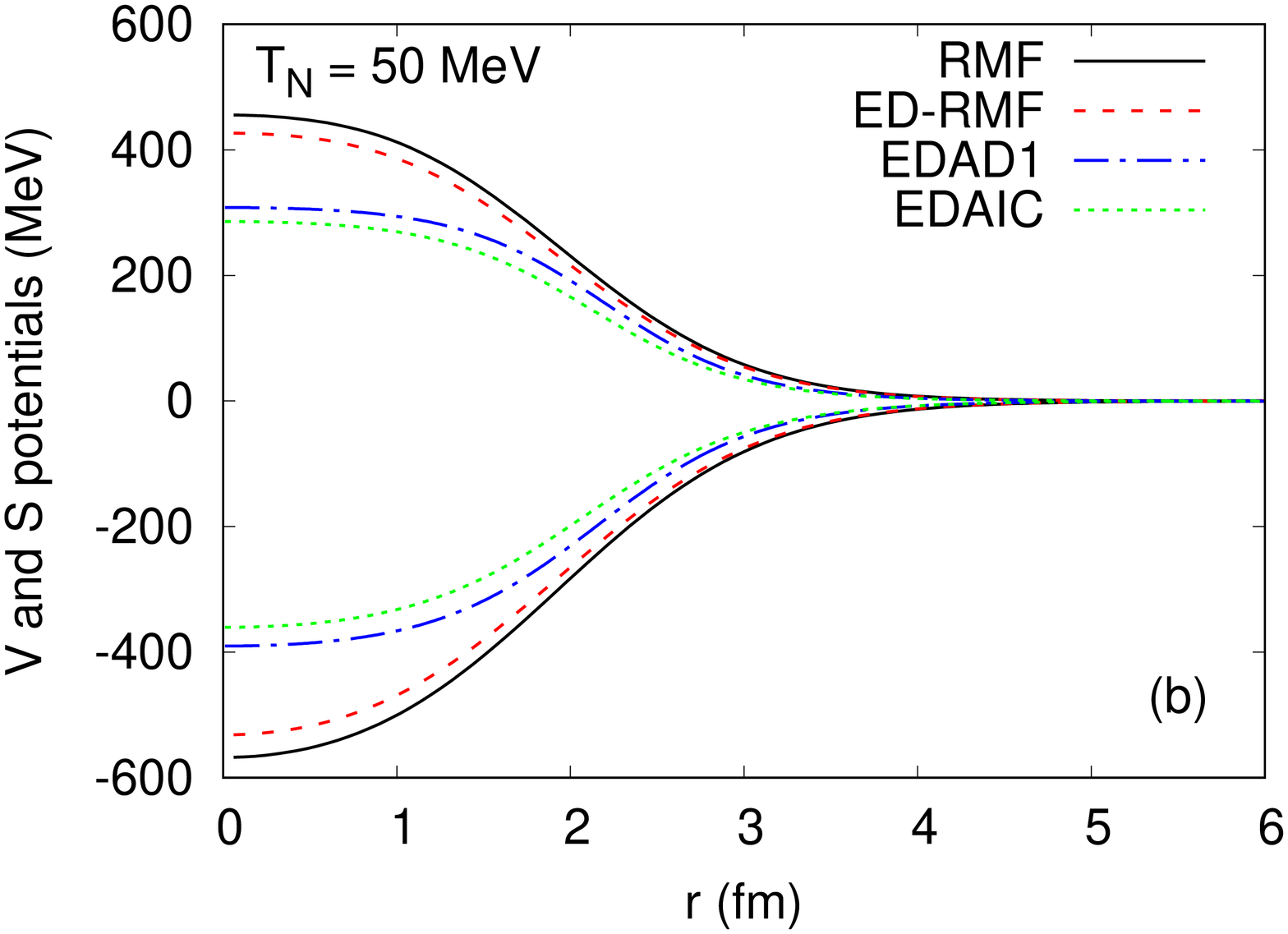}
      \includegraphics[width=.23\textwidth,angle=270]{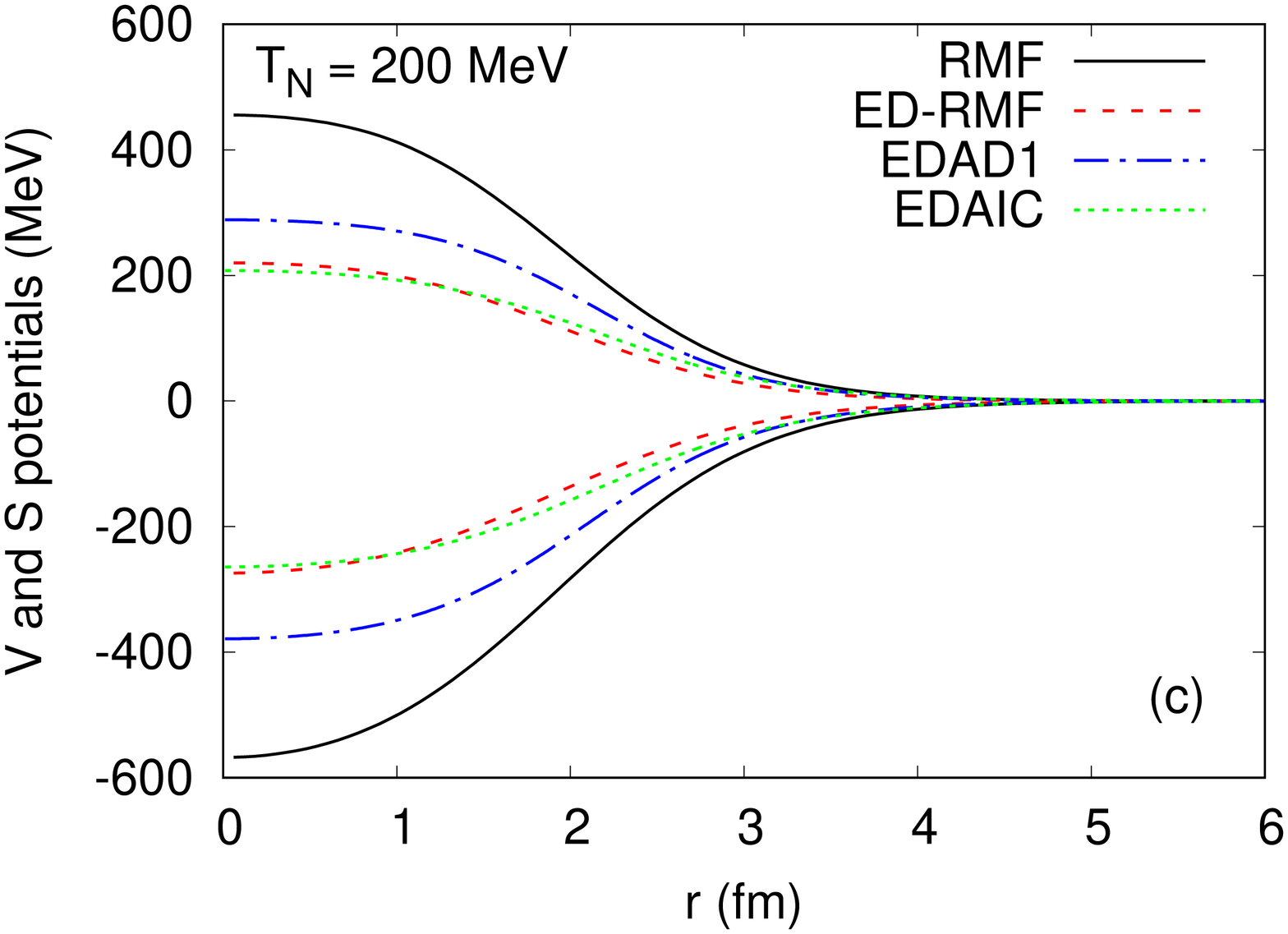}\\
      \includegraphics[width=.23\textwidth,angle=270]{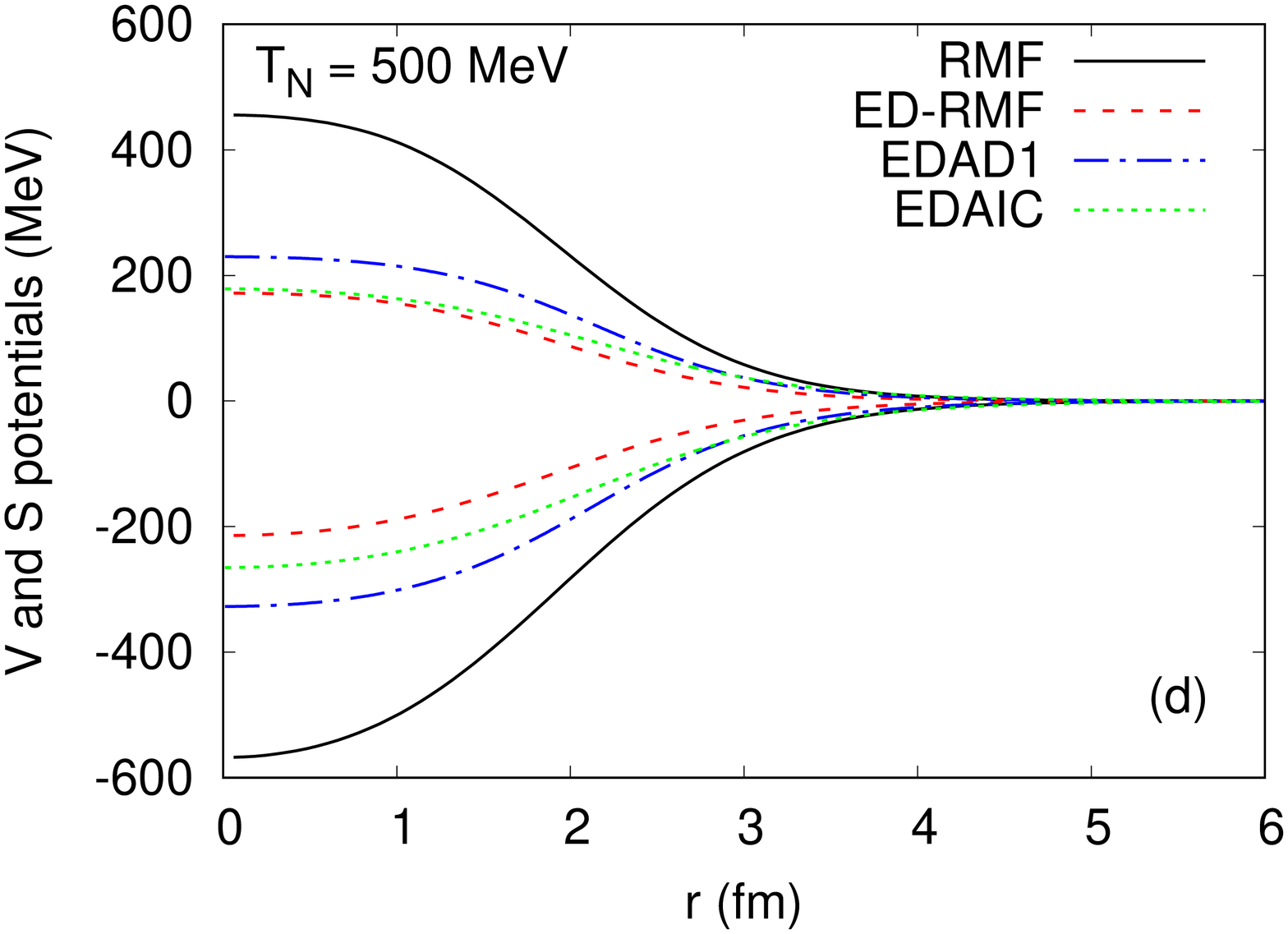}
      \includegraphics[width=.23\textwidth,angle=270]{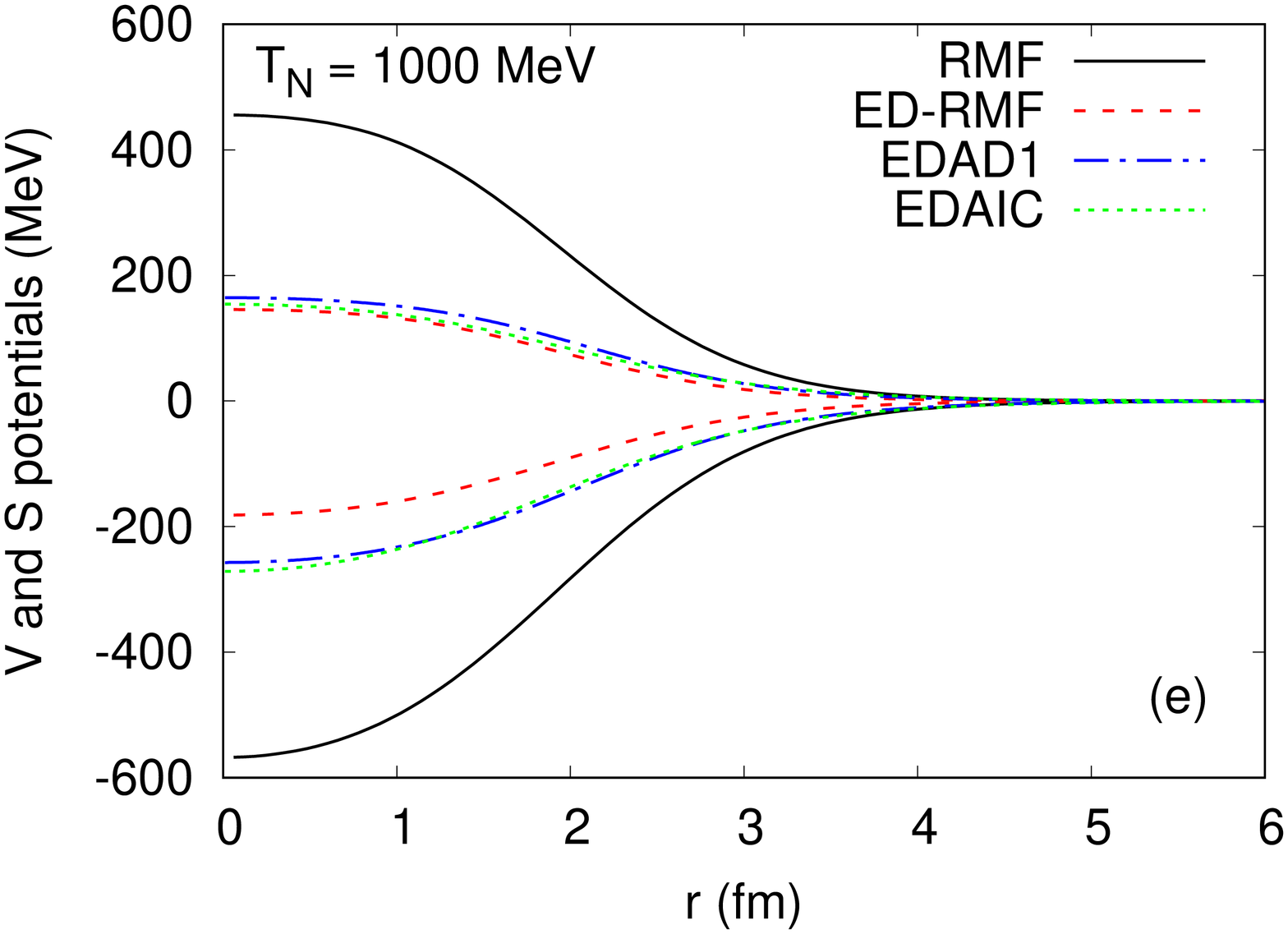}
      \includegraphics[width=.23\textwidth,angle=270]{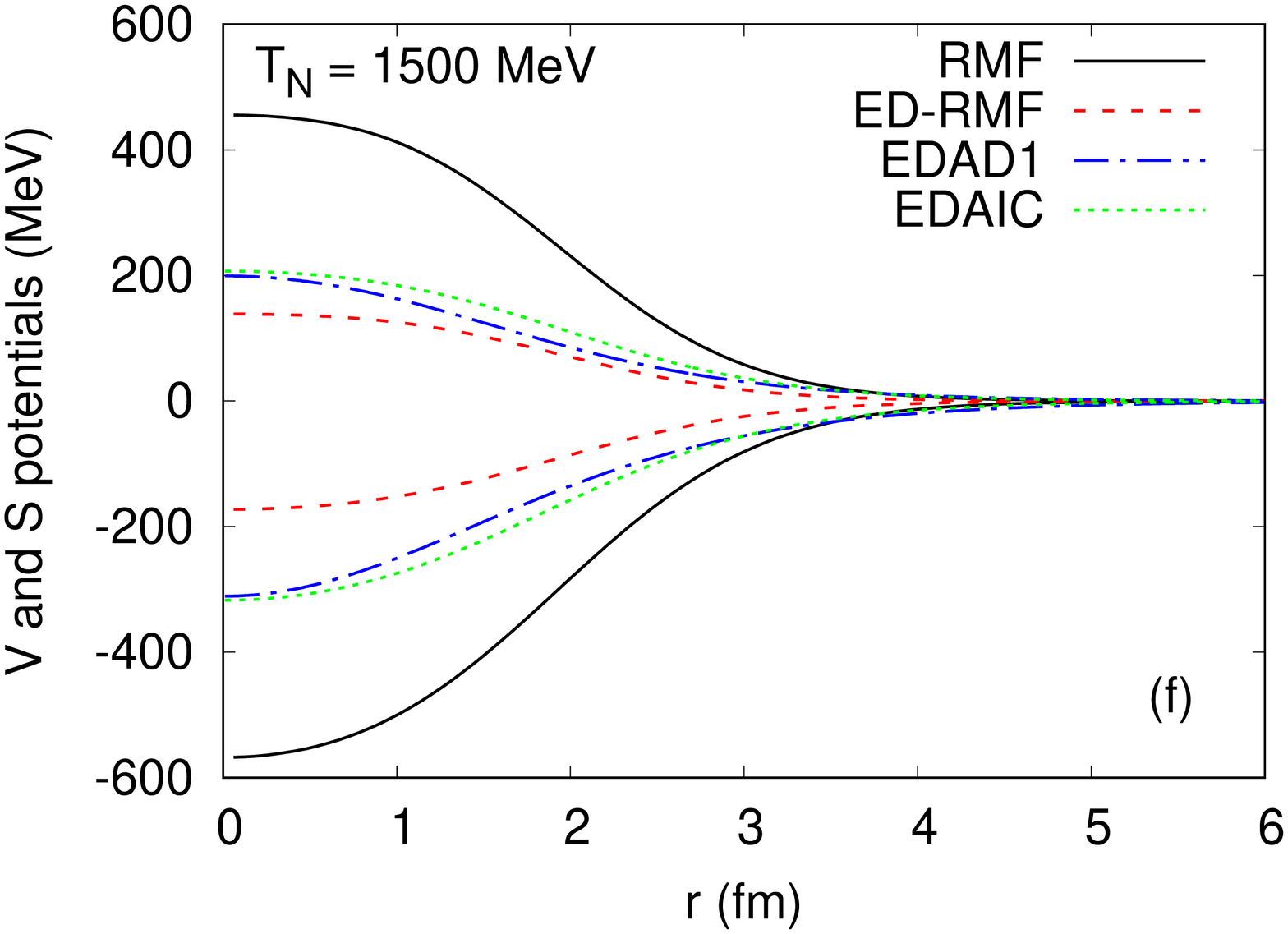}
  \caption{Vector (positive) and scalar (negative) potentials as a function of the position in the $^{12}$C nucleus. Each panel corresponds to a different kinetic energy of nucleon. 
Only the real part is represented for the EDAD1 and EDAI-C potentials.}
  \label{fig:potentials}  
\end{figure*}

In Fig.~\ref{fig:potentials} we show the scalar and vector potentials of the RMF, ED-RMF, EDAD1, and EDAI-C approaches, represented as functions of position $r$. Each panel corresponds to a different kinetic energy of the outgoing nucleon. The RMF potentials are energy-independent, the others decrease with increasing energies. 
In Figs.~\ref{fig:potentials}(a) and (b), corresponding to very small $T_p$, the ED-RMF potentials are very close to the RMF one, solving the orthogonalization problem in this energy region that is the most sensitive to it. The EDAD1 and EDAI-C, on the contrary, are considerably smaller than RMF so one should be cautious with their predictions for such kinematics. Figs.~\ref{fig:potentials}(c), (d), (e) correspond to $200<T_p<1000$ MeV. We observe that the ED-RMF, EDAD1 and EDAI-C potentials are close to each other and continuously decrease with energy. In the last panel, Fig.~\ref{fig:potentials}(f), $T_p=1500$ MeV is out of the range where the EDAD1 and EDAI-C potentials were fitted. 
Indeed, one sees that the EDAD1 and EDAI-C potentials are slightly larger than in the previous kinematics, which should be understood as a consequence of the extrapolation method.\\

In the {\it HF-CRPA model}~\cite{Pandey15,Pandey16} the bound state wave functions are obtained with a self-consistent Hartree-Fock model using an extended Skyrme force for the nucleon-nucleon interaction~\cite{Waroquier:1986mj}.
The same mean-field potential obtained for the initial state is used to compute the final-state nucleon wave functions, therefore including the essential features of orthogonality as discussed above.
Contrary to the relativistic approaches the nuclear current is obtained from the standard non-relativistic reduction of the single nucleon current as explained in~\cite{walecka04,fetterwalecka1971}.
This mean-field picture, which gives an adequate description of the genuine quasielastic cross section, is then extended with collective excitations of the nucleus in the CRPA approach. 
Although inherently non-relativistic, the calculations are effectively relativized according to the scheme of \cite{Jeschonnek}.
The HF-CRPA provides reliable results for very low momentum transfers where long-range correlations, that are not accounted for in a mean-field picture, contribute significantly to the cross section in the form of, {\it e.g.,} giant resonances~\cite{Jachowicz19}. 
This consistent treatment of the interaction from very low to moderate momentum transfers is important for neutrino-oscillation analyses that need to provide an adequate description of the electroweak interaction with nuclei over a broad region of phase space.\\

The {\it SuSAv2 model} is based on the scaling properties shown by the $(e,e')$ data and on RMF theory. 
When satisfied, the scaling property allows for the factorization of the inclusive cross section in terms of a single-nucleon elementary cross section and a scaling function, which contains all the nuclear complexity and depends on only one variable $\psi=\psi(\omega,q)$~\cite{Donnelly99a,Donnelly99b}, $\omega$ and $q$ being the energy and momentum transfer, respectively. 
The original SuSA model~\cite{Barbaro04,Amaro05a} uses only one universal scaling function extracted directly from the analysis of experimental data~\cite{Maieron02}. Although quite successful~\cite{Barbaro04,Amaro05a,Maieron09}, its simplicity does not allow one to model the complexity of the QE response with the desired accuracy, lacking for instance some strength in the transverse channel. 
The SuSAv2 model was proposed to overcome this limitation~\cite{Gonzalez-Jimenez14b}. It uses different scaling functions, extracted from RMF and RPWIA results, for the different responses. Thus, it effectively incorporates both regimes, RMF (for low and intermediate $q$) and RPWIA (for high $q$). 
This is achieved by using a ``blending'' function that introduces a linear combination of the RMF and RPWIA scaling functions. This function contains one adjustable $q$-dependent parameter that was fitted, once and for all, to reproduce $^{12}$C$(e,e')$X data in \cite{Megias16a}.

As mentioned in the Introduction, the agreement with the $(e,e')$ cross section data is the result of a delicate balance between the contributions from different channels.
In particular, the SuSAv2-MEC collaboration uses three incoherent contributions to the cross section: i) QE~\cite{Gonzalez-Jimenez14b}, ii) meson-exchange currents (MEC)~\cite{Megias15}, and iii) inelastic contributions~\cite{Megias16a}. The latter include all possible inelastic channels starting from the pion production threshold by making use of phenomenological fits of the single-nucleon inelastic structure functions. 
If one avoids the very low energy and momentum transfer region, the agreement with the inclusive data for electron and neutrino reactions is remarkable~\cite{Megias16a,Megias16b}.

In what follows we show that, for the QE response, the effective and {\it ad hoc} method of the SuSAv2 model for the transition from RMF to RPWIA by means of the blending function provides results that are very similar to those obtained with the relativistic mean-field models, solving the wave equation with the energy-dependent potentials.
This coincidence between two different and completely independent models reflects that, although the phenomenology
introduced in each model is quite different, both can provide a good description of the behavior of the nuclear system leading to very similar results for the inclusive responses.

\section{Scaling functions}\label{Scaling-functions}

\begin{figure*}[htbp]
  \centering
      \includegraphics[width=3cm,height=15cm,keepaspectratio,angle=270]{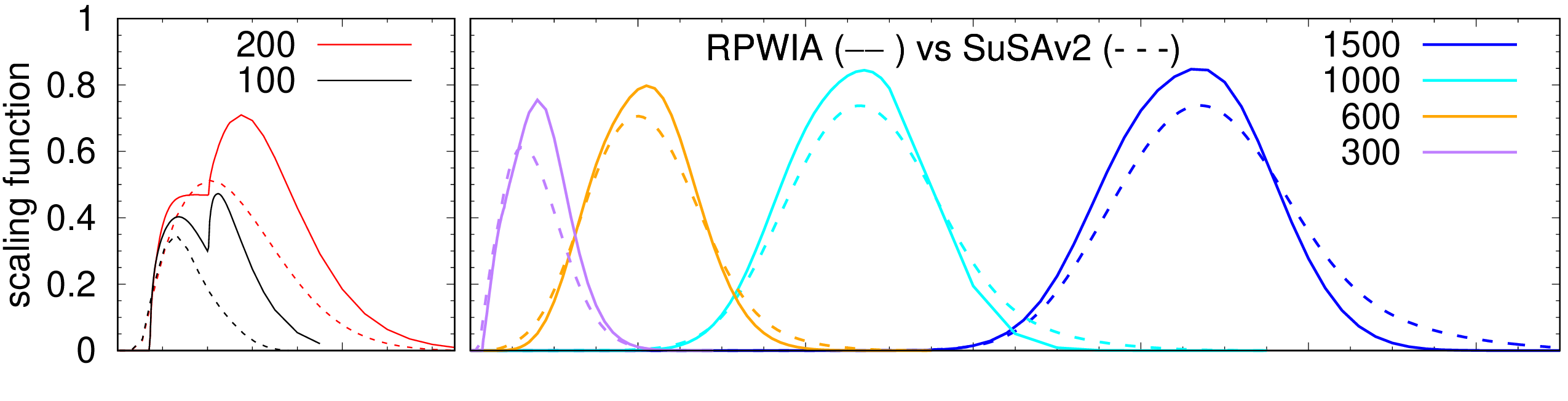}
      \includegraphics[width=3cm,height=15cm,keepaspectratio,angle=270]{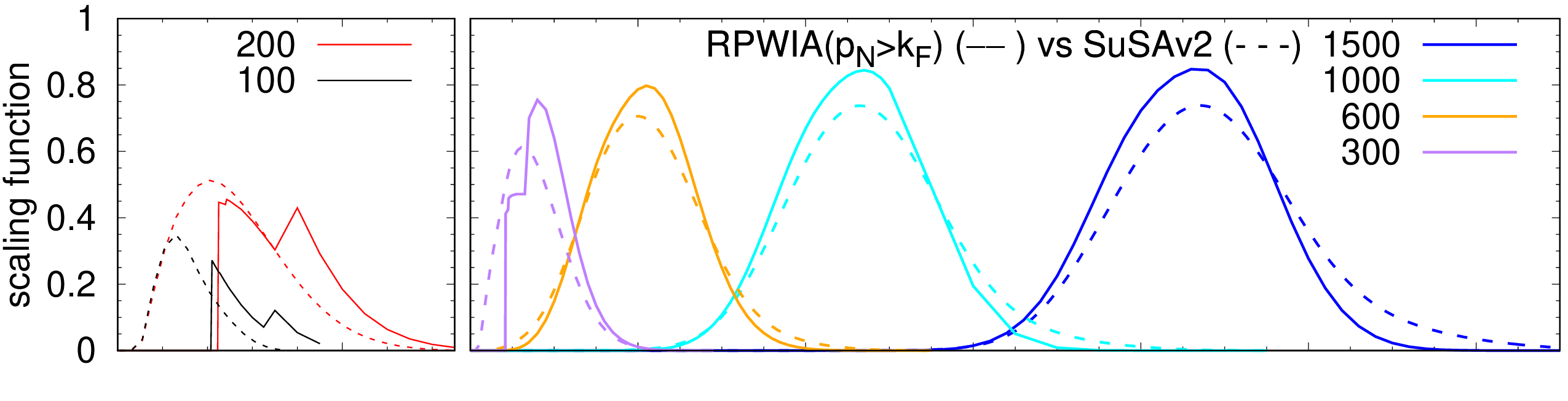}
      \includegraphics[width=3cm,height=15cm,keepaspectratio,angle=270]{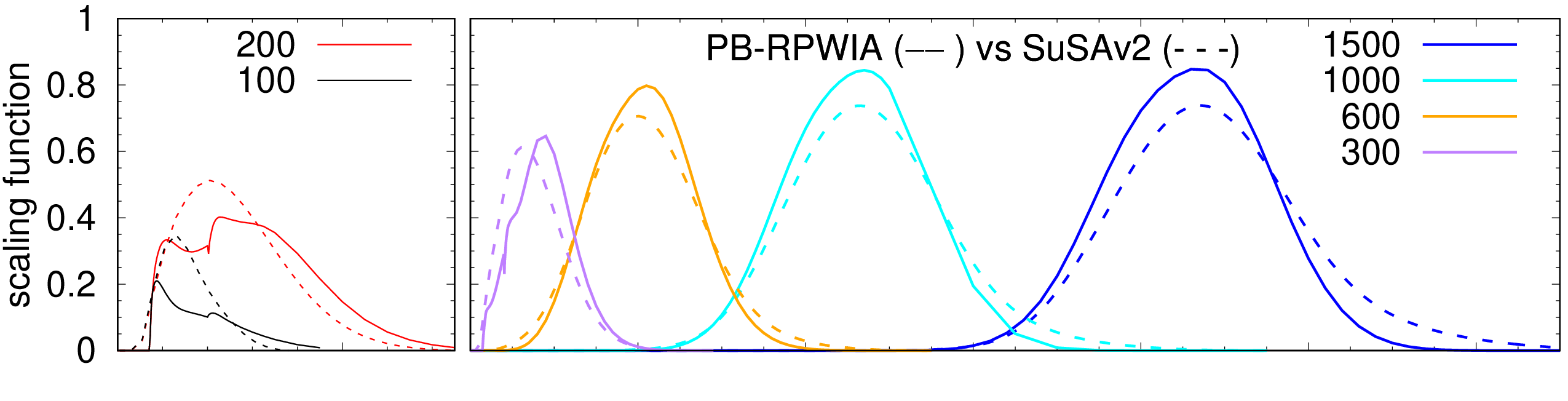}
      \includegraphics[width=3cm,height=15cm,keepaspectratio,angle=270]{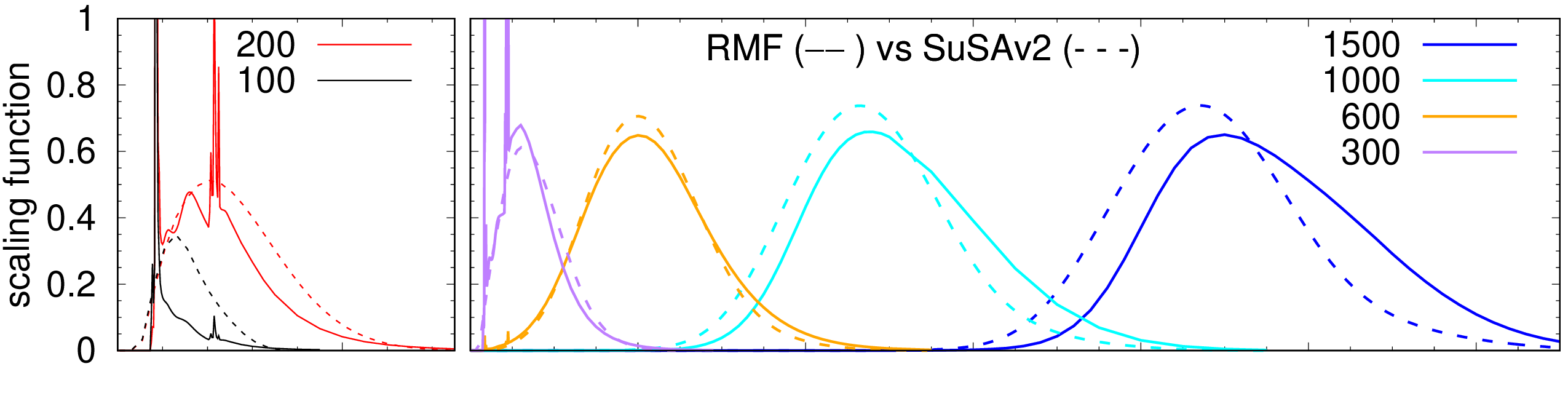}
      \includegraphics[width=3cm,height=15cm,keepaspectratio,angle=270]{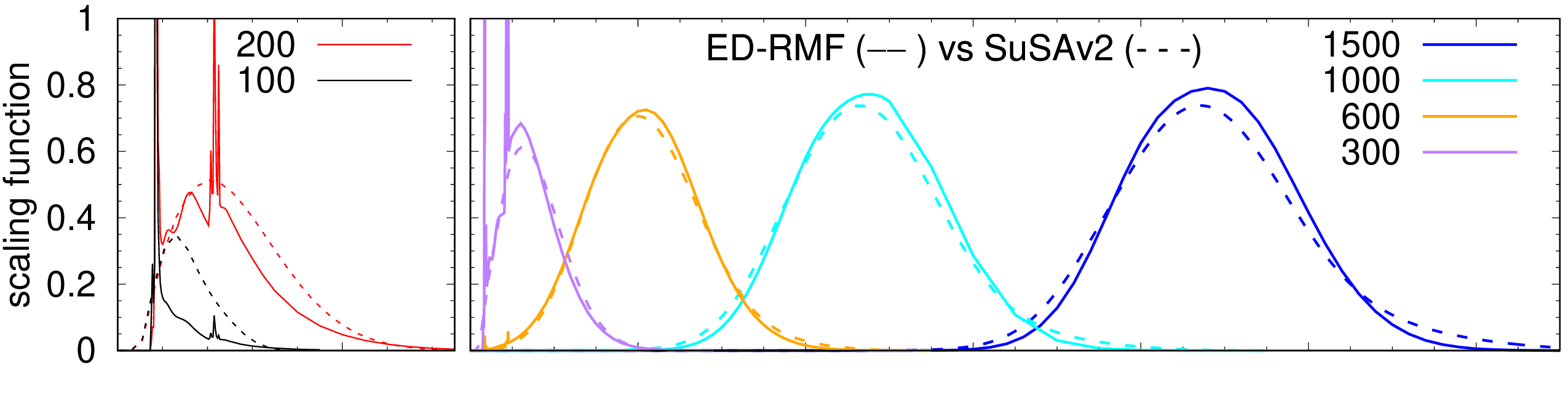}
      \includegraphics[width=3cm,height=15cm,keepaspectratio,angle=270]{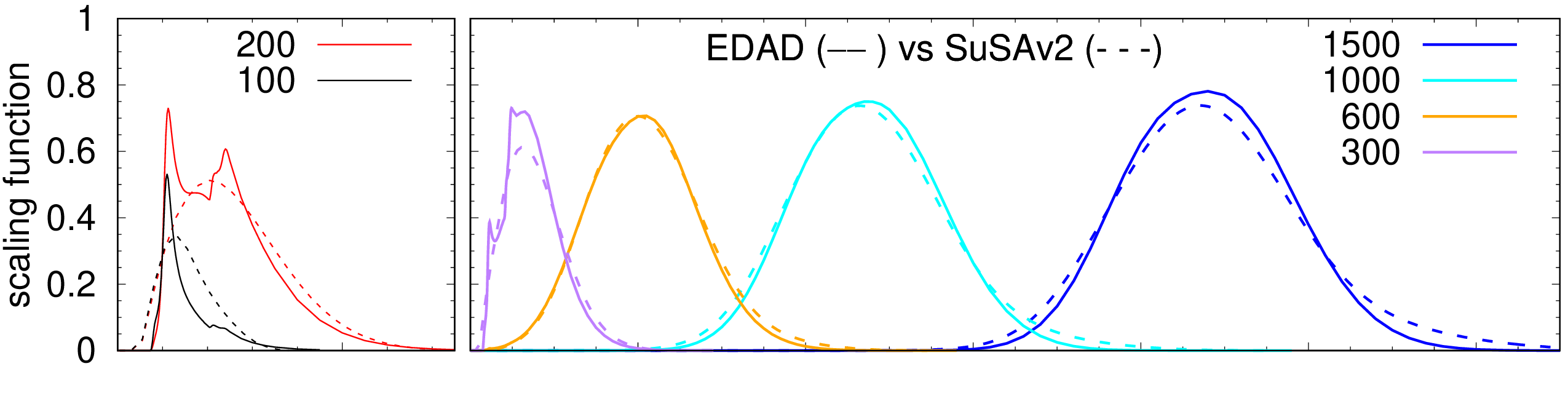}
      \includegraphics[width=3.98cm,height=15cm,keepaspectratio,angle=270]{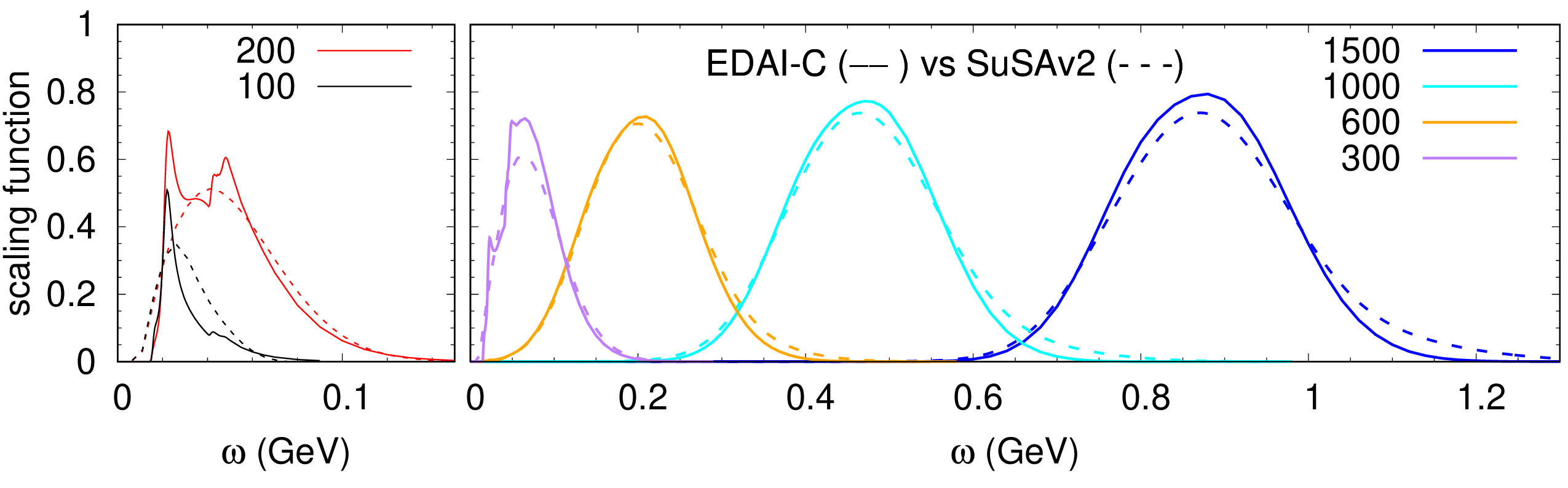}
  \caption{$^{12}$C scaling functions from SuSAv2 (dashed lines) and the relativistic mean-field models (solid lines). Each pair of curves (same color) correspond to a fixed momentum transfer $q=100$, 200, 600, 1000, and 1500 MeV. }
  \label{fig:sf-c12}
\end{figure*}

In Fig.~\ref{fig:sf-c12} we compare the scaling functions for $^{12}$C obtained with the relativistic mean-field models and the SuSAv2, described in Sect.~\ref{Models}. 
The scaling function is defined as
\ba
  f(\varepsilon_i,q,\omega) = k_F\frac{ \left[\frac{d^2\sigma}{d\omega d\cos\theta}\right]_{e,e'} }{ \sigma_{Mott}(v_L G_L + v_T G_T) }\,,\label{fscaling}
\ea
with $\left[\frac{d^2\sigma}{d\omega d\cos\theta}\right]_{e,e'}$ the inclusive cross section computed with a particular model, the denominator ($v_{L,T}$ and $G_{L,T}$ factors) is defined in~\cite{Donnelly99b}, and $k_F$ is the Fermi momentum (we use 228 MeV for carbon and 240 MeV for argon). For fixed $q$ and $\omega$ the scaling function depends very weakly on the incoming energy $\varepsilon_i$, and it has been fixed to $\varepsilon_i=3$ GeV. 
Comparing the scaling function, rather than the double differential cross section, allows us to remove most of the kinematic effects and focus on the differences arising from the nuclear modeling. 
In each panel we represent the results from one of the mean-field approaches (solid lines) and the SuSAv2 (dashed lines). For a given panel, each pair of curves (same color) correspond to a fixed $q$, which grows if one moves from left to right.

We start commenting on the region of low momentum transfer, $q\leq300$ MeV, where Pauli blocking and distortion effects are expected to be very important. 
Since these effects are fully incorporated in the RMF model, its results will be considered as the reference. 
The RPWIA scaling function is clearly too large, RPWIA($p_N>k_F$) seems to provide the right strength but the position of the distribution is clearly off, PB-RPWIA is close to RMF due to the orthogonalization procedure, but it misses the effect of the distortion. The ED-RMF is by construction essentially identical to RMF. The EDAD1 and EDAI-C results provide results close to the RMF ones. This shows that, although the initial and final states are not orthogonal, the overlap is not as large as with plane waves.
Finally, at very low $q$ one observes important differences between the SuSAv2 and the RMF model. Despite the fact that SuSAv2 incorporates the important reduction due to Pauli blocking~\cite{Megias14,Gonzalez-Jimenez14b}, it still overestimates and cannot reproduce the behavior shown by the RMF model. 
 
For $q$ values above approximately 300 MeV, where scaling is expected to work well, the results show that the scaling functions from the SuSAv2 and the energy-dependent relativistic mean-field models (ED-RMF, EDAD, and EDAI-C) notably agree. On the contrary, the RPWIA approaches yield, as expected, a scaling function which is higher than SuSAv2 at the peak and more symmetric, whereas the RMF model moves too much strength to high energies when the momentum transfer is high and the RPWIA result should be recovered: these two observations motivated the construction of the SuSAv2 model.
  
\begin{figure*}[htbp]
  \centering
      \includegraphics[width=3cm,height=15cm,keepaspectratio,angle=270]{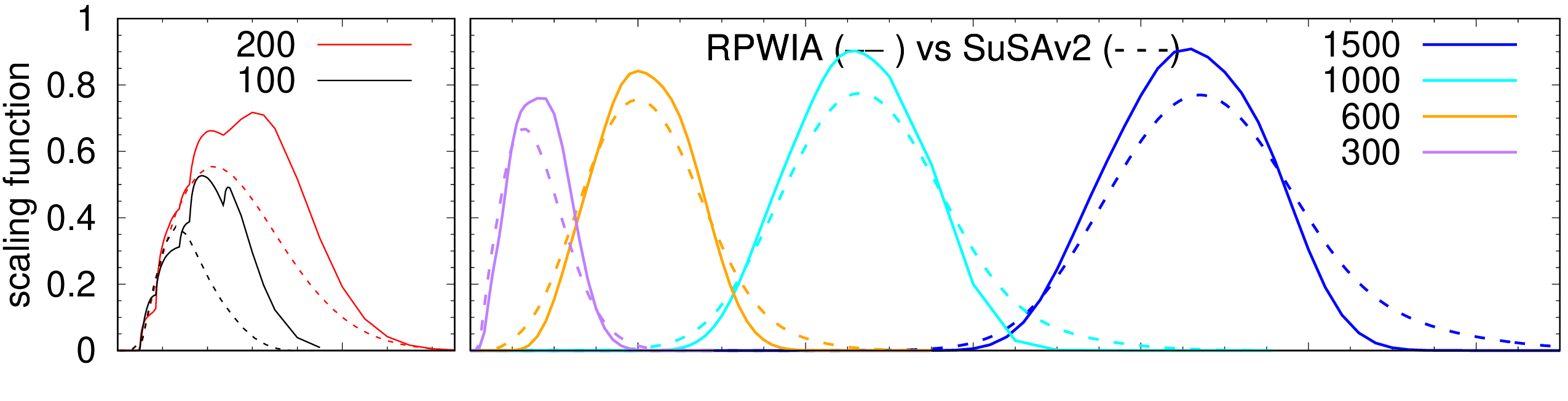}
      \includegraphics[width=3cm,height=15cm,keepaspectratio,angle=270]{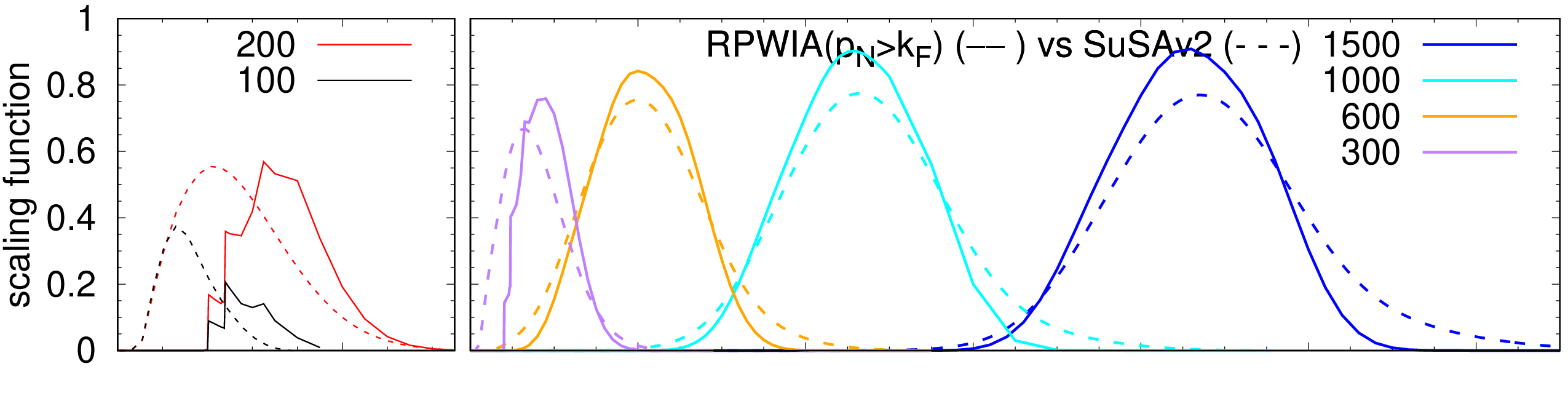}
      \includegraphics[width=3cm,height=15cm,keepaspectratio,angle=270]{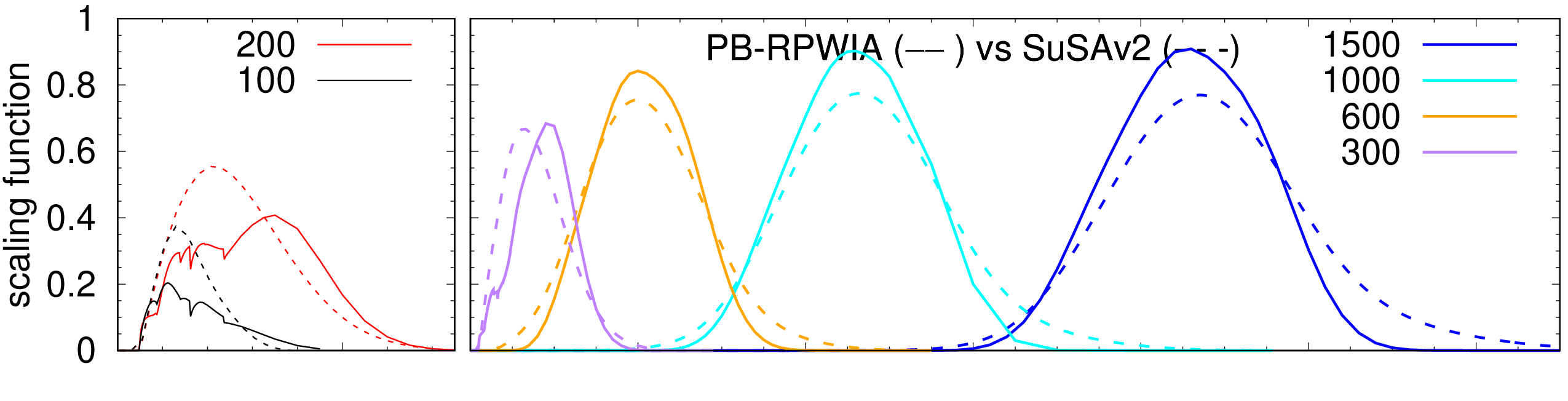}
      \includegraphics[width=3cm,height=15cm,keepaspectratio,angle=270]{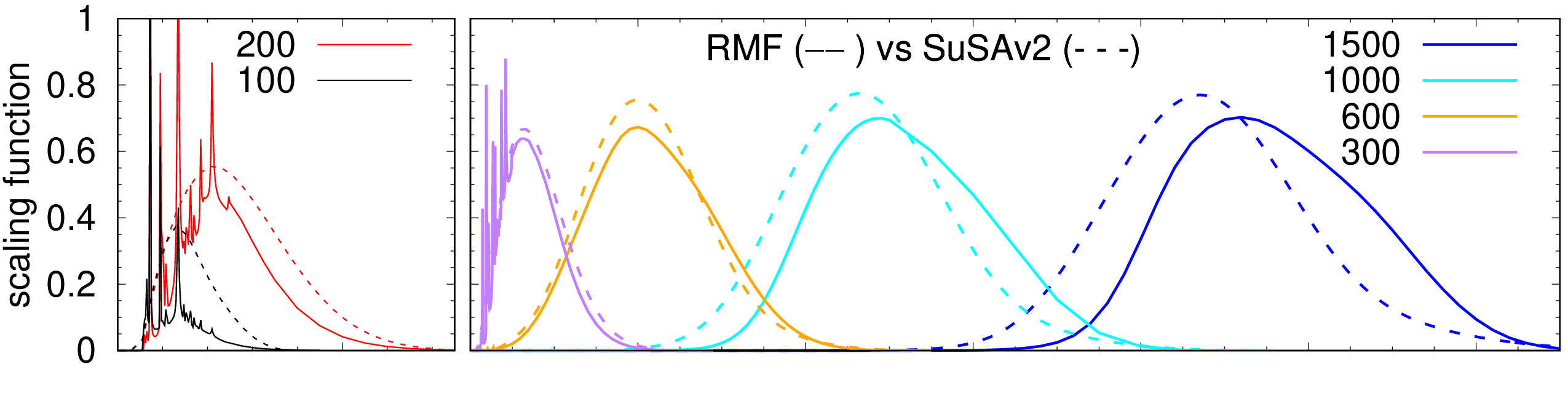}
      \includegraphics[width=3cm,height=15cm,keepaspectratio,angle=270]{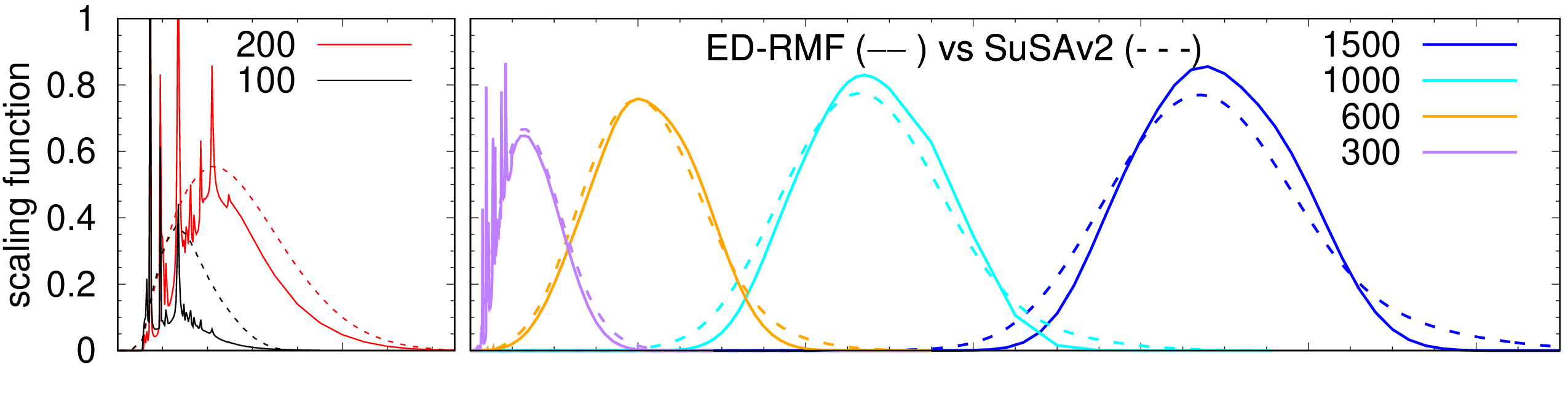}
      \includegraphics[width=3cm,height=15cm,keepaspectratio,angle=270]{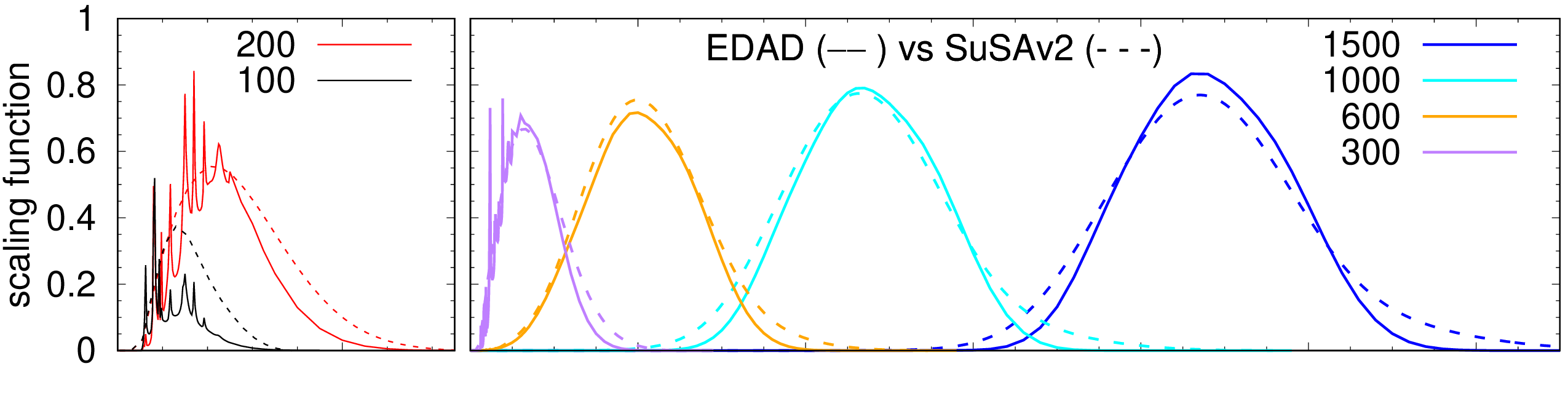}
      \includegraphics[width=3.98cm,height=15cm,keepaspectratio,angle=270]{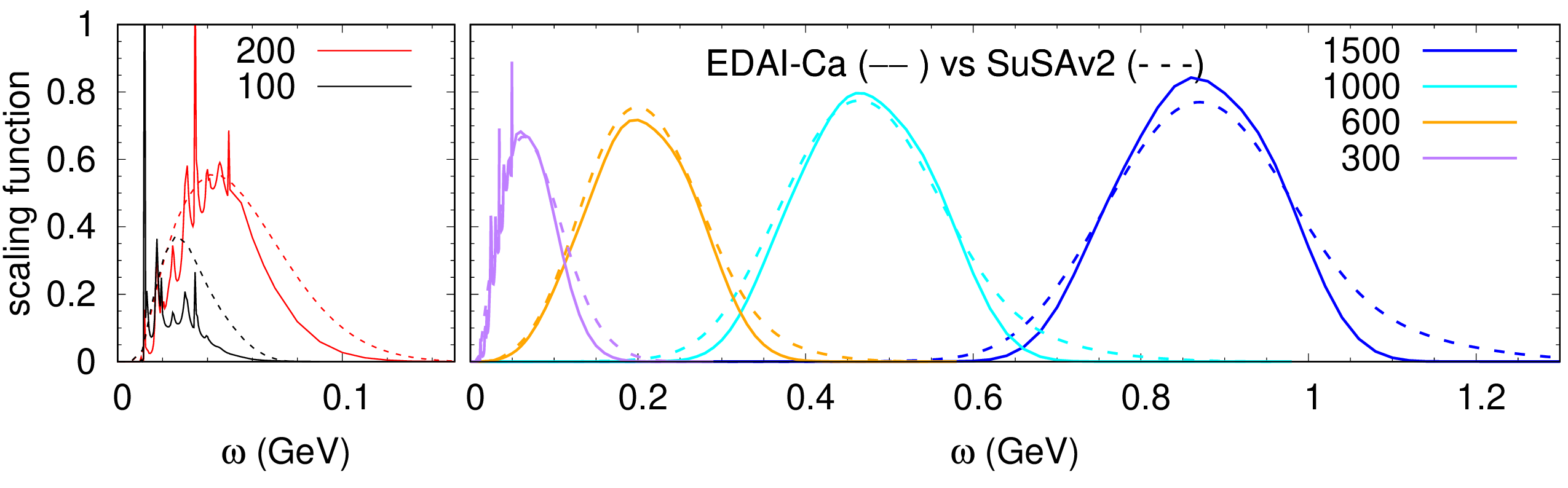}
  \caption{$^{40}$Ar scaling functions from SuSAv2 (dashed lines) and the relativistic mean-field models (solid lines). Each pair of curves (same color) correspond to a fixed momentum transfer $q=100$, 200, 600, 1000, and 1500 MeV. }
  \label{fig:sf-ar40}
\end{figure*}

Argon will be the main target in the DUNE neutrino-oscillation experiment~\cite{DUNE16}, which implies a much heavier nucleus than has typically been employed in previous experiments. Likewise, the analysis of neutrino reactions on $^{12}$C and other light nuclei, and its extrapolation to $^{40}$Ar is of special relevance for assessing design choices for the DUNE near detectors. Therefore, the results shown in this work are of paramount importance in the field to understand the neutrino-argon interaction with high precision as well as to understand the extrapolation between different nuclear targets. Since SuSAv2 was constructed from the carbon-12 RMF-RPWIA scaling functions only~\cite{Gonzalez-Jimenez14b}, its comparison with the results from the RMF-based models is an important test, that may guide future developments. 
Thus, in Fig.~\ref{fig:sf-ar40} we present the same results as in Fig.~\ref{fig:sf-c12} but for $^{40}$Ar. 
The potential EDAI-Ca, instead of EDAI-C, has been used in this case for the final-state nucleon in the EDAI caculations (last row in Fig.~\ref{fig:sf-ar40}). This is based on the idea that the mean-field potential for calcium should be similar to the one for argon.
We stress here that the RMF and ED-RMF calculations were performed for $^{40}$Ar, and that all approaches describe the initial state using the $^{40}$Ar RMF potential.
In general, the same discussions made for $^{12}$C also apply in this case.\\

\begin{figure*}[htbp]
  \centering
    \includegraphics[width=.5\textwidth,angle=270]{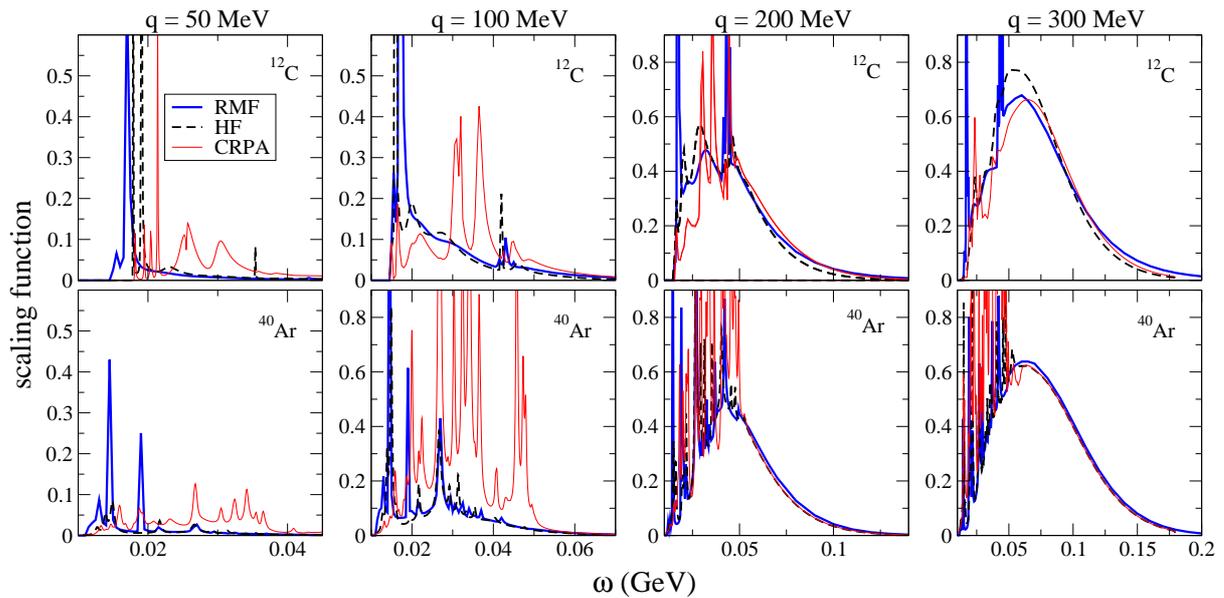}
  \caption{RMF, HF and HF-CRPA scaling functions for small $q$ values represented as a function of the energy transfer $\omega$. Upper (lower) panels correspond to carbon (argon) nucleus.}
  \label{fig:RMFvsCRPA}
\end{figure*}

At low energy and momentum transfer one expects sizeable contributions from collective nuclear effects that are beyond the pure mean field, and therefore not included in the RMF calculations. 
Some of these effects are accounted for in the HF-CRPA model.
In Fig.~\ref{fig:RMFvsCRPA} we compare the RMF, HF, and HF-CRPA results for some small values of $q$, for carbon and argon targets.
The HF and RMF approaches are very similar for the lowest $q$-values presented. 
The CRPA provides an additional enhancement of the cross section for $q$ up to around $200~\mathrm{MeV}$. 
For $q\gtrsim 300~\mathrm{MeV}$ the CRPA reduces the HF cross section, bringing the HF-CRPA and RMF results close to each other.

\section{Comparison with JLab inclusive data}\label{JLab-comparison}

\begin{figure*}[htbp]
  \centering
      \includegraphics[width=.23\textwidth,angle=270]{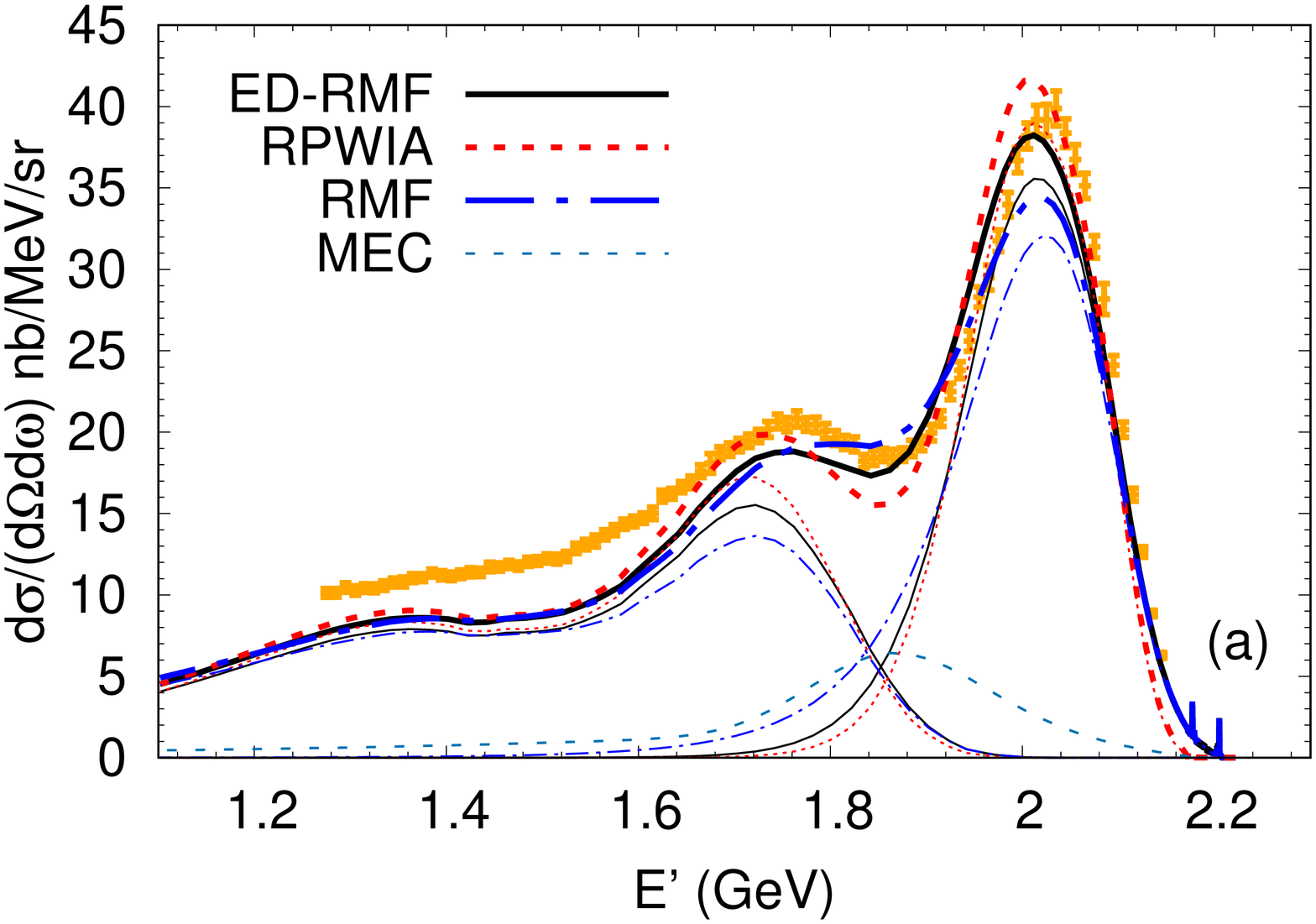}
      \includegraphics[width=.23\textwidth,angle=270]{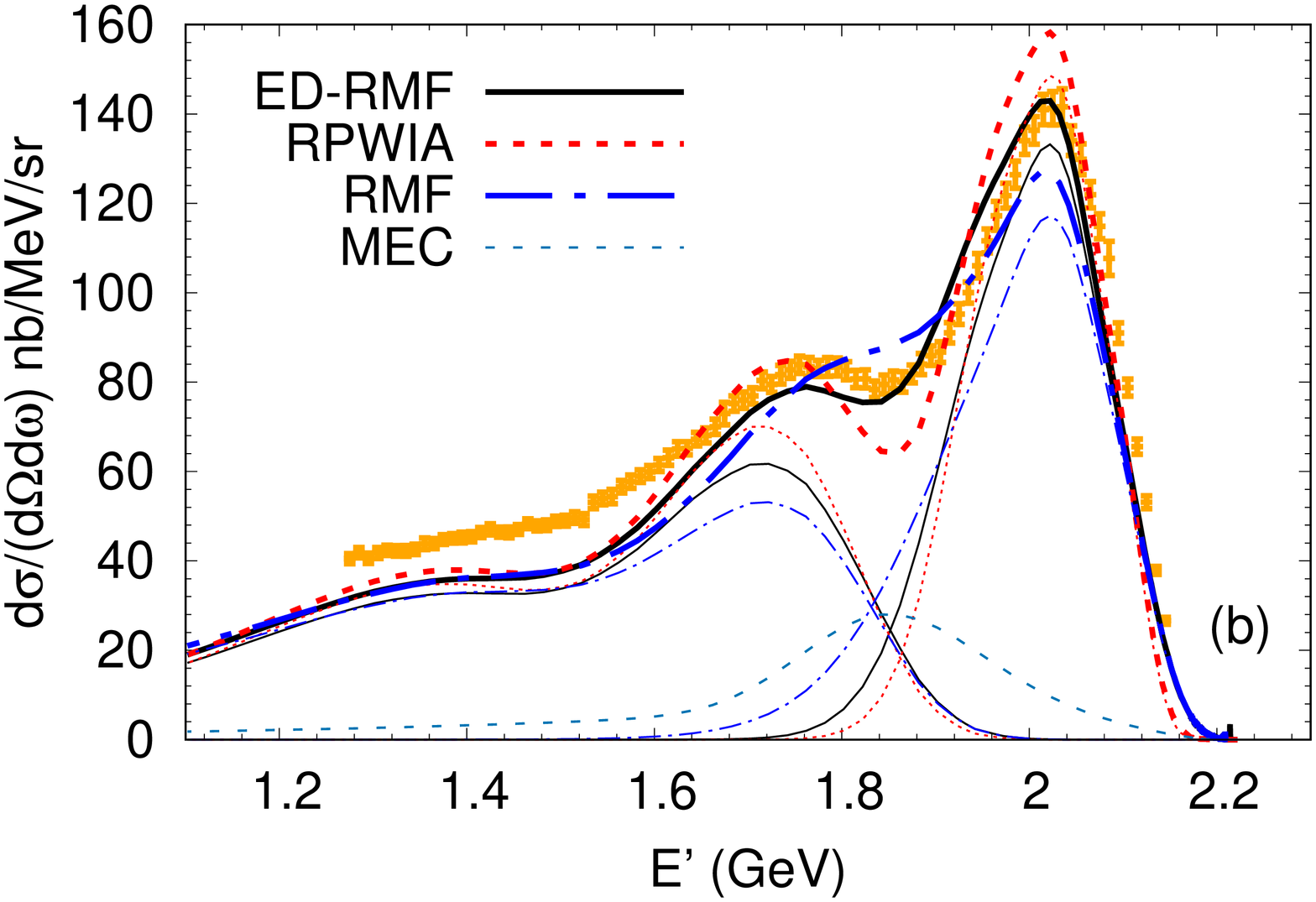}
      \includegraphics[width=.23\textwidth,angle=270]{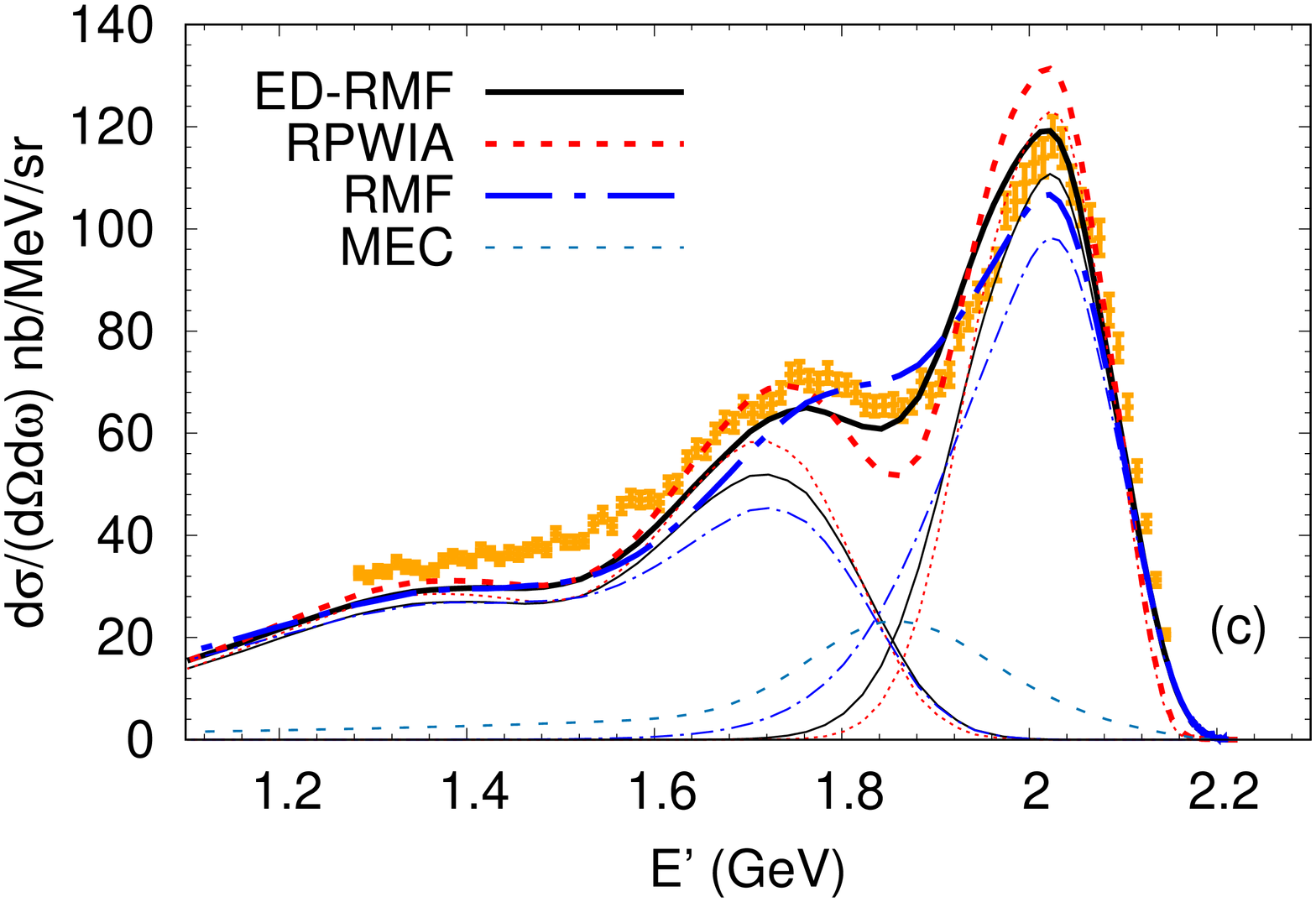}
  \caption{RPWIA, RMF and ED-RMF predictions compare with the JLab data from \cite{JLab_Ar40,JLab_Ti48} 
  ($\varepsilon_i = 2222$ MeV, $\theta_e=15.541$ deg). 
  Panels (a), (b) and (c) correspond to scattering on the target nuclei $^{12}$C, $^{48}$Ti and $^{40}$Ar, respectively. Thinner lines represent the QE, MEC and SPP contributions, thicker lines show the sum.}
  \label{fig:model-data1}
      \includegraphics[width=.23\textwidth,angle=270]{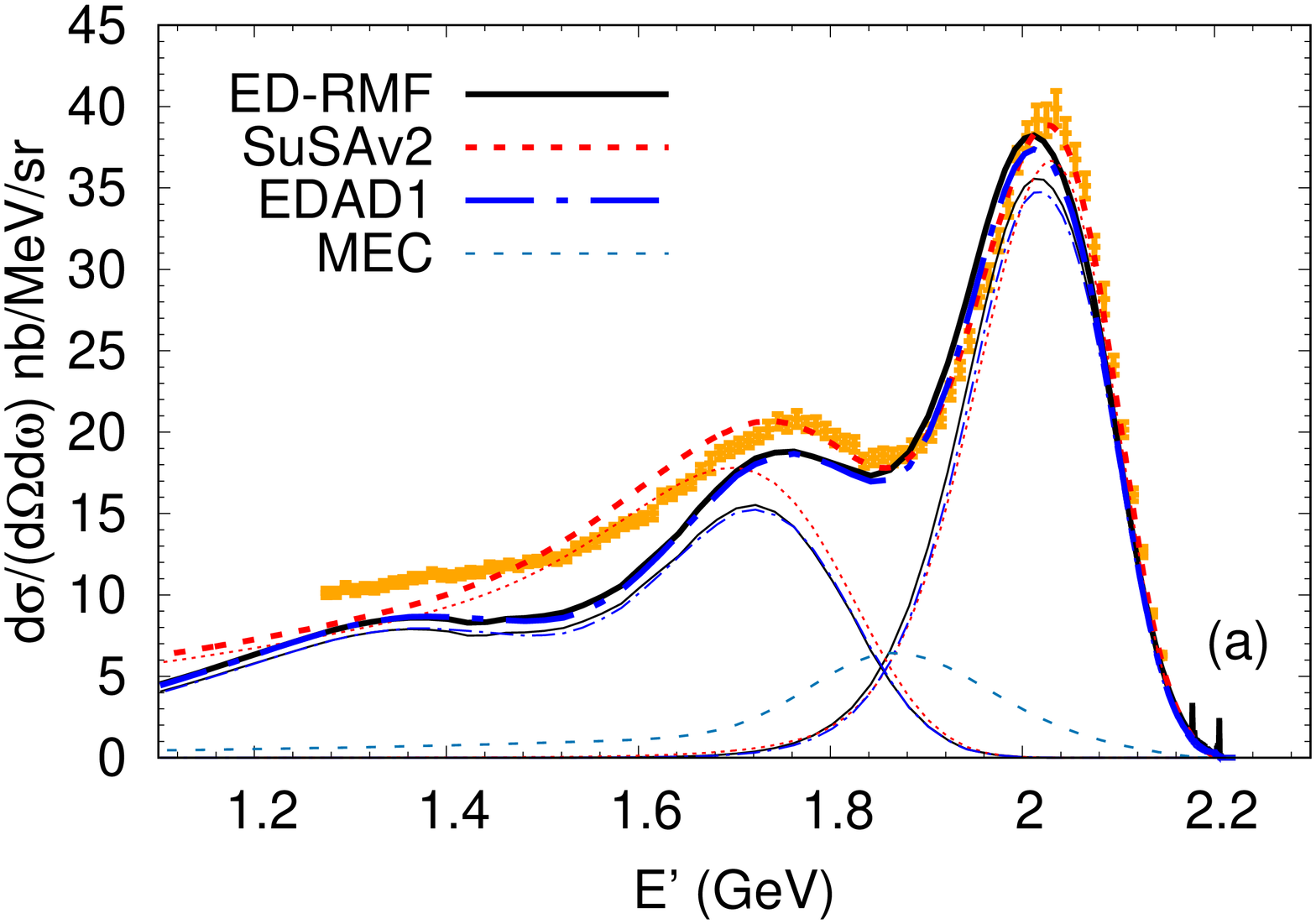}
      \includegraphics[width=.23\textwidth,angle=270]{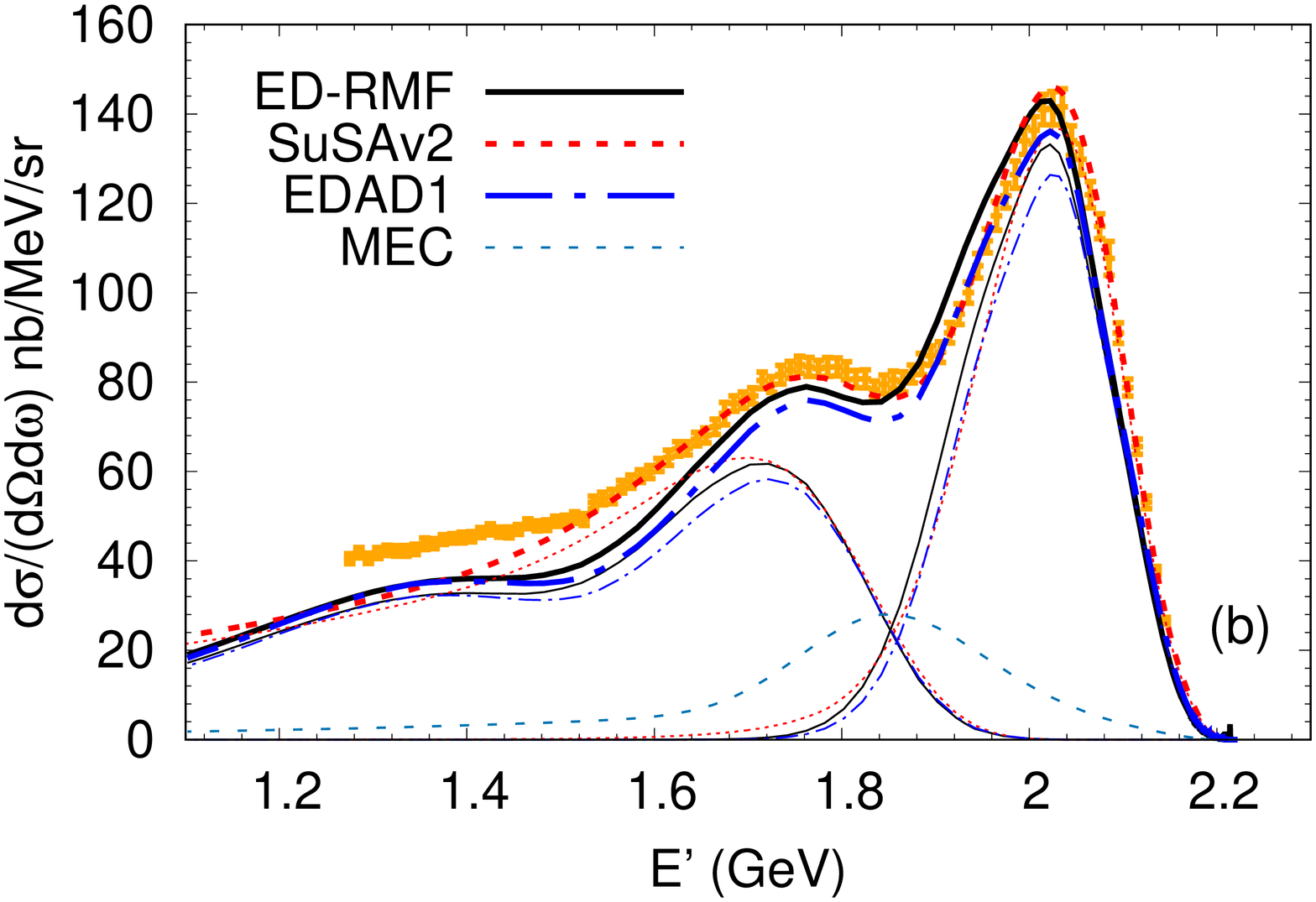}
      \includegraphics[width=.23\textwidth,angle=270]{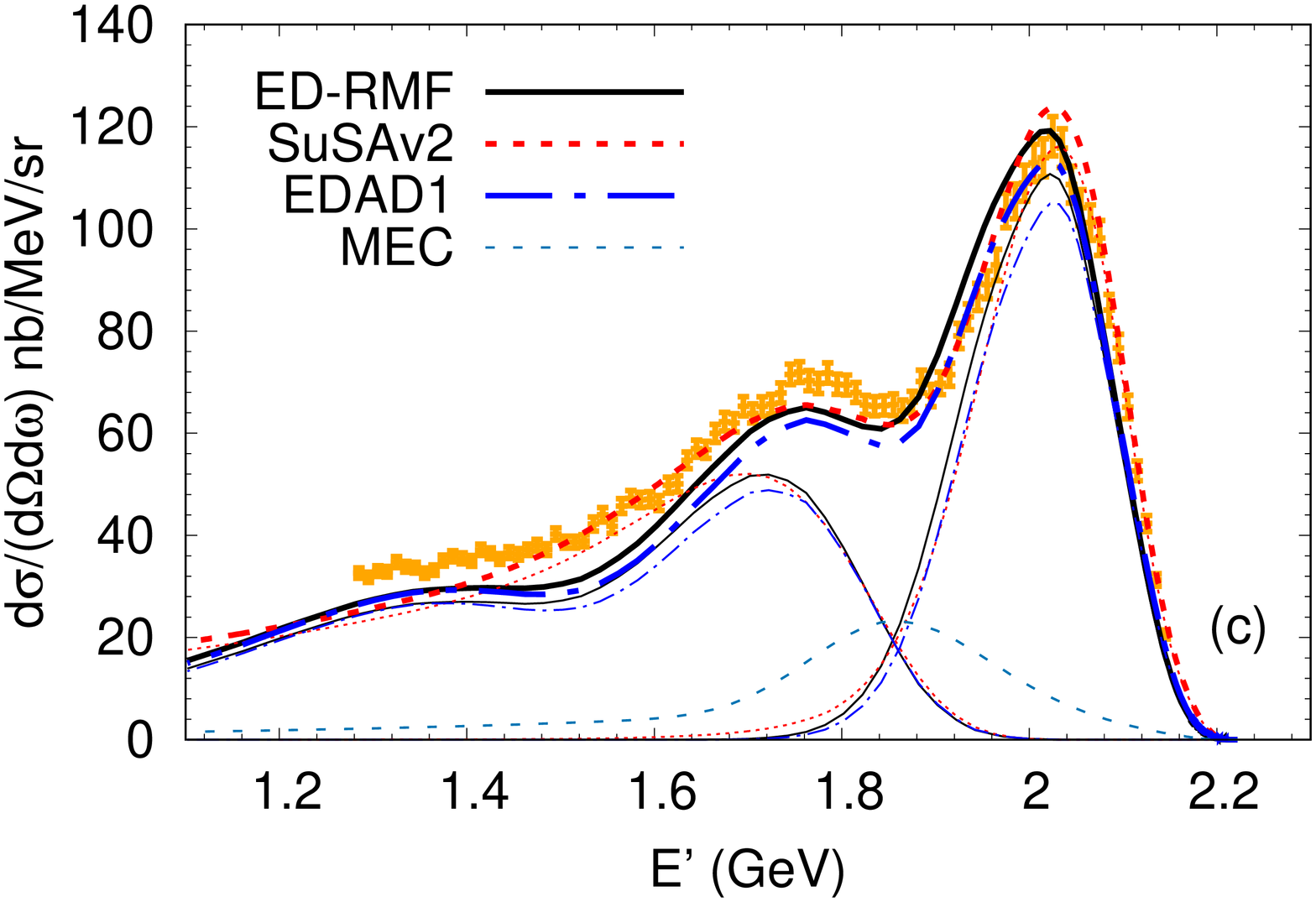}
  \caption{As Fig.~\ref{fig:model-data1} but for the SuSAv2~\cite{Barbaro19}, ED-RMF and EDAD-1 results.  }
  \label{fig:model-data2}
\end{figure*}

In Fig.~\ref{fig:model-data1} the $(e,e')$ JLab data for carbon, argon and titanium~\cite{JLab_Ar40,JLab_Ti48} are compared with the results from RPWIA, RMF, and ED-RMF models.
Although the focus of this work is on the QE response (corresponding to the peak at $E'\approx2$ GeV in Figs.~\ref{fig:model-data1}), in order to compare with these data we added more contributions.
Thus, the first bump starting from the left-hand side corresponds to single-pion production (SPP), computed with the model described in \cite{Gonzalez-Jimenez19,Gonzalez-Jimenez17}. Following the discussion in \cite{Gonzalez-Jimenez19}, we do not include medium modification of the delta width (more details can be found in \cite{Gonzalez-Jimenez18,Nikolakopoulos18a}). 
The second bump, filling the `dip' between SPP and the QE peak is the MEC contribution, taken from~\cite{Barbaro19}.

One observes that the RPWIA results clearly overestimate the QE peak and slightly the SPP one. In the dip region, however, it underpredicts the data. The opposite occurs in RMF, {\it i.e.} it underpredicts the peaks and overpredicts the dip region. This is a consequence of the redistribution of the strength from the peaks to the tails. The ED-RMF results lie in between RMF and RPWIA ones, providing notably better agreement with the experimental data.

Fig.~\ref{fig:model-data2} is analogous to Fig.~\ref{fig:model-data1} but, in this case, we compare the SuSAv2 results, previously presented in \cite{Barbaro19}, with the ED-RMF and EDAD-1 ones. 
The same MEC model has been used in the three calculations. The models for the inelastic response, however, are different. 
In the SuSAv2 approach the inclusive inelastic structure functions are modeled, thus, including all possible inelastic channels from the pion threshold. On the contrary, the ED-RMF and EDAD-1 results correspond to a microscopic calculation that accounts for the single-pion production channel only. This explains that the inelastic contribution in SuSAv2 is larger than that in the ED-RMF and EDAD-1 models.
The modeling of the QE responses is also different and has been discussed in the previous sections.

The level of agreement of SuSAv2, ED-RMF and EDAD-1 with the JLab data in the QE region is similar. However, the QE response in SuSAv2 is somewhat larger and shifted to higher $E'$ values, which seems to improve the agreement with data. A shift in the RMF-based approaches could be achieved by using more realistic values of the shell binding energies. Notice that currently, for sake of consistency, we are using the eigenvalues of the RMF hamiltonian. 
Additionally, in the case of ED-RMF model, a softer blending function $f_b$ in the region $T_N\approx200$ MeV would also lead to more RPWIA-like responses, providing results closer to SuSAv2 ones.

\section{Conclusions}\label{Conclusions}

We have studied the inclusive QE scaling functions arising from using different mean-field based models and compared with the SuSAv2 scaling approach in a large range of momentum transfer $50<q<1500$ MeV. 
By analyzing different ingredients in the models we have studied and quantified the impact of several nuclear effects such as the distortion of the outgoing nucleon, Pauli blocking and long-range nuclear correlations.

We have shown that the effective and {\it ad hoc} approach followed in the SuSAv2 model for merging RMF and RPWIA by means of a blending function is to a large extent equivalent to solving the wave equation for the scattered nucleon with relativistic energy-dependent real potentials. It is important to point out that the optical potentials EDAD1 and EDAI~\cite{Cooper93} were independently extracted by fits to elastic proton-nucleus scattering data.  
Thus, the coincidence between the outcomes of these completely different and independent approaches, namely, SuSAv2 and models using energy-dependent relativistic potentials, sets strong assurances of the capabilities of both approaches of incorporating the phenomenology to constrain the QE response.
We also found a satisfactory agreement between the energy-dependent relativistic mean-field model and the recent JLab inclusive data for carbon, argon, and titanium.

Therefore, the present work is of relevance not only for ongoing experiments but also for the next generation of neutrino-oscillation experiments (HyperKamiokande~\cite{HyperK15} and DUNE~\cite{DUNE16}) that will require a percent-level understanding of neutrino-nucleus interactions and the capability of using nuclear models to extrapolate between different nuclear targets.

\begin{acknowledgments}
This work was partially supported 
by the Spanish Ministerio de Economia y Competitividad and ERDF (European Regional Development Fund) under contracts FIS2017-88410-P, 
by the Junta de Andalucia (FQM 160, SOMM17/6105/UGR), 
by the Spanish Consolider-Ingenio 2000 program CPAN (CSD2007-0042), 
by Spanish Government (FPA2015-65035-P and RTI2018-098868-B-I00), 
by the Research Foundation Flanders (FWO-Flanders) and Special Research Fund, Ghent University, 
as well as in part by the Office of Nuclear Physics of the US Department of Energy under Grant Contract DE-FG02-94ER40818 (T. W. D.).
R.G.J. acknowledges support by Comunidad de Madrid and UCM under the contract No. 2017-T2/TIC-5252. 
M.B.B. acknowledges support by the INFN under project Iniziativa Specifica MANYBODY and the University of Turin under Projects BARM-RILO-17-01 and BARM-FFABR-17-01.
G.D.M. acknowledges support from a P2IO-CNRS grant and from CEA, CNRS/IN2P3 and P2IO, France; and 
by the European Union's Horizon 2020 research and innovation programme under the Marie Sklodowska-Curie grant agreement No 839481. 
K.N. was partially supported by NCN Opus Grant No. 2016/21/B/ST2/01092 and by the Special Research Fund, Ghent University.
The computations of this work were performed in EOLO, the HPC funded by MECD and MICINN as a contribution to CEI Moncloa; the HPC for physics funded in part by UCM and ERDF, CEI Moncloa; and the Stevin Supercomputer Infrastructure provided by Ghent University, the Hercules Foundation and the Flemish Government.
\end{acknowledgments}

\bibliographystyle{apsrev4-1}
{\small
\bibliography{bibliography}

\begin{thebibliography}{52}%
\makeatletter
\providecommand \@ifxundefined [1]{%
 \@ifx{#1\undefined}
}%
\providecommand \@ifnum [1]{%
 \ifnum #1\expandafter \@firstoftwo
 \else \expandafter \@secondoftwo
 \fi
}%
\providecommand \@ifx [1]{%
 \ifx #1\expandafter \@firstoftwo
 \else \expandafter \@secondoftwo
 \fi
}%
\providecommand \natexlab [1]{#1}%
\providecommand \enquote  [1]{``#1''}%
\providecommand \bibnamefont  [1]{#1}%
\providecommand \bibfnamefont [1]{#1}%
\providecommand \citenamefont [1]{#1}%
\providecommand \href@noop [0]{\@secondoftwo}%
\providecommand \href [0]{\begingroup \@sanitize@url \@href}%
\providecommand \@href[1]{\@@startlink{#1}\@@href}%
\providecommand \@@href[1]{\endgroup#1\@@endlink}%
\providecommand \@sanitize@url [0]{\catcode `\\12\catcode `\$12\catcode
  `\&12\catcode `\#12\catcode `\^12\catcode `\_12\catcode `\%12\relax}%
\providecommand \@@startlink[1]{}%
\providecommand \@@endlink[0]{}%
\providecommand \url  [0]{\begingroup\@sanitize@url \@url }%
\providecommand \@url [1]{\endgroup\@href {#1}{\urlprefix }}%
\providecommand \urlprefix  [0]{URL }%
\providecommand \Eprint [0]{\href }%
\providecommand \doibase [0]{http://dx.doi.org/}%
\providecommand \selectlanguage [0]{\@gobble}%
\providecommand \bibinfo  [0]{\@secondoftwo}%
\providecommand \bibfield  [0]{\@secondoftwo}%
\providecommand \translation [1]{[#1]}%
\providecommand \BibitemOpen [0]{}%
\providecommand \bibitemStop [0]{}%
\providecommand \bibitemNoStop [0]{.\EOS\space}%
\providecommand \EOS [0]{\spacefactor3000\relax}%
\providecommand \BibitemShut  [1]{\csname bibitem#1\endcsname}%
\let\auto@bib@innerbib\@empty
\bibitem [{\citenamefont {Golan}\ \emph {et~al.}(2012)\citenamefont {Golan},
  \citenamefont {Juszczak},\ and\ \citenamefont {Sobczyk}}]{Golan12}%
  \BibitemOpen
  \bibfield  {author} {\bibinfo {author} {\bibfnamefont {T.}~\bibnamefont
  {Golan}}, \bibinfo {author} {\bibfnamefont {C.}~\bibnamefont {Juszczak}}, \
  and\ \bibinfo {author} {\bibfnamefont {J.}~\bibnamefont {Sobczyk}},\ }\href
  {\doibase 10.1103/PhysRevC.86.015505} {\bibfield  {journal} {\bibinfo
  {journal} {Phys. Rev. C}\ }\textbf {\bibinfo {volume} {86}},\ \bibinfo
  {pages} {015505} (\bibinfo {year} {2012})}\BibitemShut {NoStop}%
\bibitem [{\citenamefont {Andreopoulos}\ \emph {et~al.}(2010)\citenamefont
  {Andreopoulos} \emph {et~al.}}]{Andreopoulos10}%
  \BibitemOpen
  \bibfield  {author} {\bibinfo {author} {\bibfnamefont {C.}~\bibnamefont
  {Andreopoulos}} \emph {et~al.},\ }\href {\doibase 10.1016/j.nima.2009.12.009}
  {\bibfield  {journal} {\bibinfo  {journal} {Nucl. Instrum. Meth. Phys. Res.
  Sect. A}\ }\textbf {\bibinfo {volume} {614}},\ \bibinfo {pages} {87}
  (\bibinfo {year} {2010})}\BibitemShut {NoStop}%
\bibitem [{\citenamefont {Hayato}(2009)}]{Hayato09}%
  \BibitemOpen
  \bibfield  {author} {\bibinfo {author} {\bibfnamefont {Y.}~\bibnamefont
  {Hayato}},\ }\href@noop {} {\bibfield  {journal} {\bibinfo  {journal} {Acta
  Phys. Pol. B}\ }\textbf {\bibinfo {volume} {40}},\ \bibinfo {pages} {2477}
  (\bibinfo {year} {2009})}\BibitemShut {NoStop}%
\bibitem [{\citenamefont {Buss}\ \emph {et~al.}(2012)\citenamefont {Buss},
  \citenamefont {Gaitanos}, \citenamefont {Gallmeister}, \citenamefont {van
  Hees}, \citenamefont {Kaskulov}, \citenamefont {Lalakulich}, \citenamefont
  {Larionov}, \citenamefont {Leitner}, \citenamefont {Weil},\ and\
  \citenamefont {Mosel}}]{Buss12}%
  \BibitemOpen
  \bibfield  {author} {\bibinfo {author} {\bibfnamefont {O.}~\bibnamefont
  {Buss}}, \bibinfo {author} {\bibfnamefont {T.}~\bibnamefont {Gaitanos}},
  \bibinfo {author} {\bibfnamefont {K.}~\bibnamefont {Gallmeister}}, \bibinfo
  {author} {\bibfnamefont {H.}~\bibnamefont {van Hees}}, \bibinfo {author}
  {\bibfnamefont {M.}~\bibnamefont {Kaskulov}}, \bibinfo {author}
  {\bibfnamefont {O.}~\bibnamefont {Lalakulich}}, \bibinfo {author}
  {\bibfnamefont {A.}~\bibnamefont {Larionov}}, \bibinfo {author}
  {\bibfnamefont {T.}~\bibnamefont {Leitner}}, \bibinfo {author} {\bibfnamefont
  {J.}~\bibnamefont {Weil}}, \ and\ \bibinfo {author} {\bibfnamefont
  {U.}~\bibnamefont {Mosel}},\ }\href {\doibase 10.1016/j.physrep.2011.12.001}
  {\bibfield  {journal} {\bibinfo  {journal} {Physics Reports}\ }\textbf
  {\bibinfo {volume} {512}},\ \bibinfo {pages} {1 } (\bibinfo {year} {2012})},\
  \bibinfo {note} {transport-theoretical Description of Nuclear
  Reactions}\BibitemShut {NoStop}%
\bibitem [{\citenamefont {Abe}\ \emph {et~al.}(2015)\citenamefont {Abe} \emph
  {et~al.}}]{HyperK15}%
  \BibitemOpen
  \bibfield  {author} {\bibinfo {author} {\bibfnamefont {K.}~\bibnamefont
  {Abe}} \emph {et~al.},\ }\href {\doibase 10.1093/ptep/ptv061} {\bibfield
  {journal} {\bibinfo  {journal} {PTEP}\ }\textbf {\bibinfo {volume} {2015}},\
  \bibinfo {pages} {053C02} (\bibinfo {year} {2015})}\BibitemShut {NoStop}%
\bibitem [{\citenamefont {Donnelly}\ \emph {et~al.}(2017)\citenamefont
  {Donnelly}, \citenamefont {Formaggio}, \citenamefont {Holstein},
  \citenamefont {Milner},\ and\ \citenamefont {Surrow}}]{Foundations17}%
  \BibitemOpen
  \bibfield  {author} {\bibinfo {author} {\bibfnamefont {T.~W.}\ \bibnamefont
  {Donnelly}}, \bibinfo {author} {\bibfnamefont {J.~A.}\ \bibnamefont
  {Formaggio}}, \bibinfo {author} {\bibfnamefont {B.~R.}\ \bibnamefont
  {Holstein}}, \bibinfo {author} {\bibfnamefont {R.~G.}\ \bibnamefont
  {Milner}}, \ and\ \bibinfo {author} {\bibfnamefont {B.}~\bibnamefont
  {Surrow}},\ }\enquote {\bibinfo {title} {Introduction to lepton
  scattering},}\ in\ \href {\doibase 10.1017/9781139028264.008} {\emph
  {\bibinfo {booktitle} {Foundations of Nuclear and Particle Physics}}}\
  (\bibinfo  {publisher} {Cambridge University Press},\ \bibinfo {year}
  {2017})\ pp.\ \bibinfo {pages} {151--187}\BibitemShut {NoStop}%
\bibitem [{\citenamefont {Alvarez-Ruso}\ \emph {et~al.}(2018)\citenamefont
  {Alvarez-Ruso}, \citenamefont {Sajjad Athar}, \citenamefont {Barbaro},
  \citenamefont {Cherdack}, \citenamefont {Christy}, \citenamefont {Coloma},
  \citenamefont {Donnelly}, \citenamefont {Dytman}, \citenamefont
  {de Gouvêa}, \citenamefont {Hill}, \citenamefont {Huber}, \citenamefont
  {Jachowicz}, \citenamefont {Katori}, \citenamefont {Kronfeld}, \citenamefont
  {Mahn}, \citenamefont {Martini}, \citenamefont {Morf\'in}, \citenamefont
  {Nieves}, \citenamefont {Perdue}, \citenamefont {Petti}, \citenamefont
  {Richards}, \citenamefont {S\'anchez}, \citenamefont {Sato}, \citenamefont
  {Sobczyk},\ and\ \citenamefont {Zeller}}]{Alvarez-Ruso18}%
  \BibitemOpen
  \bibfield  {author} {\bibinfo {author} {\bibfnamefont {L.}~\bibnamefont
  {Alvarez-Ruso}}, \bibinfo {author} {\bibfnamefont {M.}~\bibnamefont
  {Sajjad Athar}}, \bibinfo {author} {\bibfnamefont {M.}~\bibnamefont
  {Barbaro}}, \bibinfo {author} {\bibfnamefont {D.}~\bibnamefont {Cherdack}},
  \bibinfo {author} {\bibfnamefont {M.}~\bibnamefont {Christy}}, \bibinfo
  {author} {\bibfnamefont {P.}~\bibnamefont {Coloma}}, \bibinfo {author}
  {\bibfnamefont {T.}~\bibnamefont {Donnelly}}, \bibinfo {author}
  {\bibfnamefont {S.}~\bibnamefont {Dytman}}, \bibinfo {author} {\bibfnamefont
  {A.}~\bibnamefont {de Gouvêa}}, \bibinfo {author} {\bibfnamefont
  {R.}~\bibnamefont {Hill}}, \bibinfo {author} {\bibfnamefont {P.}~\bibnamefont
  {Huber}}, \bibinfo {author} {\bibfnamefont {N.}~\bibnamefont {Jachowicz}},
  \bibinfo {author} {\bibfnamefont {T.}~\bibnamefont {Katori}}, \bibinfo
  {author} {\bibfnamefont {A.}~\bibnamefont {Kronfeld}}, \bibinfo {author}
  {\bibfnamefont {K.}~\bibnamefont {Mahn}}, \bibinfo {author} {\bibfnamefont
  {M.}~\bibnamefont {Martini}}, \bibinfo {author} {\bibfnamefont
  {J.}~\bibnamefont {Morf\'in}}, \bibinfo {author} {\bibfnamefont
  {J.}~\bibnamefont {Nieves}}, \bibinfo {author} {\bibfnamefont
  {G.}~\bibnamefont {Perdue}}, \bibinfo {author} {\bibfnamefont
  {R.}~\bibnamefont {Petti}}, \bibinfo {author} {\bibfnamefont
  {D.}~\bibnamefont {Richards}}, \bibinfo {author} {\bibfnamefont
  {F.}~\bibnamefont {S\'anchez}}, \bibinfo {author} {\bibfnamefont
  {T.}~\bibnamefont {Sato}}, \bibinfo {author} {\bibfnamefont {J.}~\bibnamefont
  {Sobczyk}}, \ and\ \bibinfo {author} {\bibfnamefont {G.}~\bibnamefont
  {Zeller}},\ }\href {\doibase https://doi.org/10.1016/j.ppnp.2018.01.006}
  {\bibfield  {journal} {\bibinfo  {journal} {Progress in Particle and Nuclear
  Physics}\ }\textbf {\bibinfo {volume} {100}},\ \bibinfo {pages} {1 }
  (\bibinfo {year} {2018})}\BibitemShut {NoStop}%
\bibitem [{\citenamefont {Capuzzi}\ \emph {et~al.}(1991)\citenamefont
  {Capuzzi}, \citenamefont {Giusti},\ and\ \citenamefont {Pacati}}]{Capuzzi91}%
  \BibitemOpen
  \bibfield  {author} {\bibinfo {author} {\bibfnamefont {F.}~\bibnamefont
  {Capuzzi}}, \bibinfo {author} {\bibfnamefont {C.}~\bibnamefont {Giusti}}, \
  and\ \bibinfo {author} {\bibfnamefont {F.}~\bibnamefont {Pacati}},\ }\href
  {\doibase https://doi.org/10.1016/0375-9474(91)90269-C} {\bibfield  {journal}
  {\bibinfo  {journal} {Nuclear Physics A}\ }\textbf {\bibinfo {volume}
  {524}},\ \bibinfo {pages} {681 } (\bibinfo {year} {1991})}\BibitemShut
  {NoStop}%
\bibitem [{\citenamefont {Meucci}\ \emph {et~al.}(2009)\citenamefont {Meucci},
  \citenamefont {Caballero}, \citenamefont {Giusti}, \citenamefont {Pacati},\
  and\ \citenamefont {Ud\'{\i}as}}]{Meucci09}%
  \BibitemOpen
  \bibfield  {author} {\bibinfo {author} {\bibfnamefont {A.}~\bibnamefont
  {Meucci}}, \bibinfo {author} {\bibfnamefont {J.~A.}\ \bibnamefont
  {Caballero}}, \bibinfo {author} {\bibfnamefont {C.}~\bibnamefont {Giusti}},
  \bibinfo {author} {\bibfnamefont {F.~D.}\ \bibnamefont {Pacati}}, \ and\
  \bibinfo {author} {\bibfnamefont {J.~M.}\ \bibnamefont {Ud\'{\i}as}},\ }\href
  {\doibase 10.1103/PhysRevC.80.024605} {\bibfield  {journal} {\bibinfo
  {journal} {Phys. Rev. C}\ }\textbf {\bibinfo {volume} {80}},\ \bibinfo
  {pages} {024605} (\bibinfo {year} {2009})}\BibitemShut {NoStop}%
\bibitem [{\citenamefont {Ivanov}\ \emph
  {et~al.}(2016{\natexlab{a}})\citenamefont {Ivanov}, \citenamefont {Vignote},
  \citenamefont {\'Alvarez-Rodr\'{\i}guez}, \citenamefont {Meucci},
  \citenamefont {Giusti},\ and\ \citenamefont {Ud\'{\i}as}}]{Ivanov16b}%
  \BibitemOpen
  \bibfield  {author} {\bibinfo {author} {\bibfnamefont {M.~V.}\ \bibnamefont
  {Ivanov}}, \bibinfo {author} {\bibfnamefont {J.~R.}\ \bibnamefont {Vignote}},
  \bibinfo {author} {\bibfnamefont {R.}~\bibnamefont
  {\'Alvarez-Rodr\'{\i}guez}}, \bibinfo {author} {\bibfnamefont
  {A.}~\bibnamefont {Meucci}}, \bibinfo {author} {\bibfnamefont
  {C.}~\bibnamefont {Giusti}}, \ and\ \bibinfo {author} {\bibfnamefont {J.~M.}\
  \bibnamefont {Ud\'{\i}as}},\ }\href {\doibase 10.1103/PhysRevC.94.014608}
  {\bibfield  {journal} {\bibinfo  {journal} {Phys. Rev. C}\ }\textbf {\bibinfo
  {volume} {94}},\ \bibinfo {pages} {014608} (\bibinfo {year}
  {2016}{\natexlab{a}})}\BibitemShut {NoStop}%
\bibitem [{\citenamefont {Ankowski}\ \emph {et~al.}(2015)\citenamefont
  {Ankowski}, \citenamefont {Benhar},\ and\ \citenamefont
  {Sakuda}}]{Ankowski15a}%
  \BibitemOpen
  \bibfield  {author} {\bibinfo {author} {\bibfnamefont {A.~M.}\ \bibnamefont
  {Ankowski}}, \bibinfo {author} {\bibfnamefont {O.}~\bibnamefont {Benhar}}, \
  and\ \bibinfo {author} {\bibfnamefont {M.}~\bibnamefont {Sakuda}},\ }\href
  {\doibase 10.1103/PhysRevD.91.033005} {\bibfield  {journal} {\bibinfo
  {journal} {Phys. Rev. D}\ }\textbf {\bibinfo {volume} {91}},\ \bibinfo
  {pages} {033005} (\bibinfo {year} {2015})}\BibitemShut {NoStop}%
\bibitem [{\citenamefont {Ivanov}\ \emph {et~al.}(2019)\citenamefont {Ivanov},
  \citenamefont {Antonov}, \citenamefont {Megias}, \citenamefont {Caballero},
  \citenamefont {Barbaro}, \citenamefont {Amaro}, \citenamefont {Ruiz~Simo},
  \citenamefont {Donnelly},\ and\ \citenamefont {Ud\'{\i}as}}]{Ivanov19}%
  \BibitemOpen
  \bibfield  {author} {\bibinfo {author} {\bibfnamefont {M.~V.}\ \bibnamefont
  {Ivanov}}, \bibinfo {author} {\bibfnamefont {A.~N.}\ \bibnamefont {Antonov}},
  \bibinfo {author} {\bibfnamefont {G.~D.}\ \bibnamefont {Megias}}, \bibinfo
  {author} {\bibfnamefont {J.~A.}\ \bibnamefont {Caballero}}, \bibinfo {author}
  {\bibfnamefont {M.~B.}\ \bibnamefont {Barbaro}}, \bibinfo {author}
  {\bibfnamefont {J.~E.}\ \bibnamefont {Amaro}}, \bibinfo {author}
  {\bibfnamefont {I.}~\bibnamefont {Ruiz~Simo}}, \bibinfo {author}
  {\bibfnamefont {T.~W.}\ \bibnamefont {Donnelly}}, \ and\ \bibinfo {author}
  {\bibfnamefont {J.~M.}\ \bibnamefont {Ud\'{\i}as}},\ }\href {\doibase
  10.1103/PhysRevC.99.014610} {\bibfield  {journal} {\bibinfo  {journal} {Phys.
  Rev. C}\ }\textbf {\bibinfo {volume} {99}},\ \bibinfo {pages} {014610}
  (\bibinfo {year} {2019})}\BibitemShut {NoStop}%
\bibitem [{\citenamefont {Nieves}\ \emph {et~al.}(2011)\citenamefont {Nieves},
  \citenamefont {Simo},\ and\ \citenamefont {Vacas}}]{Nieves11}%
  \BibitemOpen
  \bibfield  {author} {\bibinfo {author} {\bibfnamefont {J.}~\bibnamefont
  {Nieves}}, \bibinfo {author} {\bibfnamefont {I.~R.}\ \bibnamefont {Simo}}, \
  and\ \bibinfo {author} {\bibfnamefont {M.~J.~V.}\ \bibnamefont {Vacas}},\
  }\href {\doibase 10.1103/PhysRevC.83.045501} {\bibfield  {journal} {\bibinfo
  {journal} {Phys. Rev. C}\ }\textbf {\bibinfo {volume} {83}},\ \bibinfo
  {pages} {045501} (\bibinfo {year} {2011})}\BibitemShut {NoStop}%
\bibitem [{\citenamefont {Martini}\ \emph {et~al.}(2010)\citenamefont
  {Martini}, \citenamefont {Ericson}, \citenamefont {Chanfray},\ and\
  \citenamefont {Marteau}}]{Martini10}%
  \BibitemOpen
  \bibfield  {author} {\bibinfo {author} {\bibfnamefont {M.}~\bibnamefont
  {Martini}}, \bibinfo {author} {\bibfnamefont {M.}~\bibnamefont {Ericson}},
  \bibinfo {author} {\bibfnamefont {G.}~\bibnamefont {Chanfray}}, \ and\
  \bibinfo {author} {\bibfnamefont {J.}~\bibnamefont {Marteau}},\ }\href
  {\doibase 10.1103/PhysRevC.81.045502} {\bibfield  {journal} {\bibinfo
  {journal} {Phys. Rev. C}\ }\textbf {\bibinfo {volume} {81}},\ \bibinfo
  {pages} {045502} (\bibinfo {year} {2010})}\BibitemShut {NoStop}%
\bibitem [{\citenamefont {Martini}\ \emph {et~al.}(2011)\citenamefont
  {Martini}, \citenamefont {Ericson},\ and\ \citenamefont
  {Chanfray}}]{Martini11}%
  \BibitemOpen
  \bibfield  {author} {\bibinfo {author} {\bibfnamefont {M.}~\bibnamefont
  {Martini}}, \bibinfo {author} {\bibfnamefont {M.}~\bibnamefont {Ericson}}, \
  and\ \bibinfo {author} {\bibfnamefont {G.}~\bibnamefont {Chanfray}},\ }\href
  {\doibase 10.1103/PhysRevC.84.055502} {\bibfield  {journal} {\bibinfo
  {journal} {Phys. Rev. C}\ }\textbf {\bibinfo {volume} {84}},\ \bibinfo
  {pages} {055502} (\bibinfo {year} {2011})}\BibitemShut {NoStop}%
\bibitem [{\citenamefont {Rocco}\ \emph {et~al.}(2017)\citenamefont {Rocco},
  \citenamefont {Alvarez-Ruso}, \citenamefont {Lovato},\ and\ \citenamefont
  {Nieves}}]{Rocco17}%
  \BibitemOpen
  \bibfield  {author} {\bibinfo {author} {\bibfnamefont {N.}~\bibnamefont
  {Rocco}}, \bibinfo {author} {\bibfnamefont {L.}~\bibnamefont {Alvarez-Ruso}},
  \bibinfo {author} {\bibfnamefont {A.}~\bibnamefont {Lovato}}, \ and\ \bibinfo
  {author} {\bibfnamefont {J.}~\bibnamefont {Nieves}},\ }\href {\doibase
  10.1103/PhysRevC.96.015504} {\bibfield  {journal} {\bibinfo  {journal} {Phys.
  Rev. C}\ }\textbf {\bibinfo {volume} {96}},\ \bibinfo {pages} {015504}
  (\bibinfo {year} {2017})}\BibitemShut {NoStop}%
\bibitem [{\citenamefont {Amaro}\ \emph {et~al.}(2005)\citenamefont {Amaro},
  \citenamefont {Barbaro}, \citenamefont {Caballero}, \citenamefont {Donnelly},
  \citenamefont {Molinari},\ and\ \citenamefont {Sick}}]{Amaro05a}%
  \BibitemOpen
  \bibfield  {author} {\bibinfo {author} {\bibfnamefont {J.~E.}\ \bibnamefont
  {Amaro}}, \bibinfo {author} {\bibfnamefont {M.~B.}\ \bibnamefont {Barbaro}},
  \bibinfo {author} {\bibfnamefont {J.~A.}\ \bibnamefont {Caballero}}, \bibinfo
  {author} {\bibfnamefont {T.~W.}\ \bibnamefont {Donnelly}}, \bibinfo {author}
  {\bibfnamefont {A.}~\bibnamefont {Molinari}}, \ and\ \bibinfo {author}
  {\bibfnamefont {I.}~\bibnamefont {Sick}},\ }\href {\doibase
  10.1103/PhysRevC.71.015501} {\bibfield  {journal} {\bibinfo  {journal} {Phys.
  Rev. C}\ }\textbf {\bibinfo {volume} {71}},\ \bibinfo {pages} {015501}
  (\bibinfo {year} {2005})}\BibitemShut {NoStop}%
\bibitem [{\citenamefont {Gonz\'alez-Jim\'enez}\ \emph
  {et~al.}(2014)\citenamefont {Gonz\'alez-Jim\'enez}, \citenamefont {Megias},
  \citenamefont {Barbaro}, \citenamefont {Caballero},\ and\ \citenamefont
  {Donnelly}}]{Gonzalez-Jimenez14b}%
  \BibitemOpen
  \bibfield  {author} {\bibinfo {author} {\bibfnamefont {R.}~\bibnamefont
  {Gonz\'alez-Jim\'enez}}, \bibinfo {author} {\bibfnamefont {G.~D.}\
  \bibnamefont {Megias}}, \bibinfo {author} {\bibfnamefont {M.~B.}\
  \bibnamefont {Barbaro}}, \bibinfo {author} {\bibfnamefont {J.~A.}\
  \bibnamefont {Caballero}}, \ and\ \bibinfo {author} {\bibfnamefont {T.~W.}\
  \bibnamefont {Donnelly}},\ }\href {\doibase 10.1103/PhysRevC.90.035501}
  {\bibfield  {journal} {\bibinfo  {journal} {Phys. Rev. C}\ }\textbf {\bibinfo
  {volume} {90}},\ \bibinfo {pages} {035501} (\bibinfo {year}
  {2014})}\BibitemShut {NoStop}%
\bibitem [{\citenamefont {Ivanov}\ \emph
  {et~al.}(2016{\natexlab{b}})\citenamefont {Ivanov}, \citenamefont {Megias},
  \citenamefont {Gonz{\'{a}}lez-Jim{\'{e}}nez}, \citenamefont {Moreno},
  \citenamefont {Barbaro}, \citenamefont {Caballero},\ and\ \citenamefont
  {Donnelly}}]{Ivanov16a}%
  \BibitemOpen
  \bibfield  {author} {\bibinfo {author} {\bibfnamefont {M.~V.}\ \bibnamefont
  {Ivanov}}, \bibinfo {author} {\bibfnamefont {G.~D.}\ \bibnamefont {Megias}},
  \bibinfo {author} {\bibfnamefont {R.}~\bibnamefont
  {Gonz{\'{a}}lez-Jim{\'{e}}nez}}, \bibinfo {author} {\bibfnamefont
  {O.}~\bibnamefont {Moreno}}, \bibinfo {author} {\bibfnamefont {M.~B.}\
  \bibnamefont {Barbaro}}, \bibinfo {author} {\bibfnamefont {J.~A.}\
  \bibnamefont {Caballero}}, \ and\ \bibinfo {author} {\bibfnamefont {T.~W.}\
  \bibnamefont {Donnelly}},\ }\href {\doibase 10.1088/0954-3899/43/4/045101}
  {\bibfield  {journal} {\bibinfo  {journal} {J. Phys. G: Nucl. Part. Phys.}\
  }\textbf {\bibinfo {volume} {43}},\ \bibinfo {pages} {045101} (\bibinfo
  {year} {2016}{\natexlab{b}})}\BibitemShut {NoStop}%
\bibitem [{\citenamefont {Amaro}\ \emph {et~al.}(2018)\citenamefont {Amaro},
  \citenamefont {Martinez-Consentino}, \citenamefont {Ruiz~Arriola},\ and\
  \citenamefont {Ruiz~Simo}}]{Amaro18}%
  \BibitemOpen
  \bibfield  {author} {\bibinfo {author} {\bibfnamefont {J.~E.}\ \bibnamefont
  {Amaro}}, \bibinfo {author} {\bibfnamefont {V.~L.}\ \bibnamefont
  {Martinez-Consentino}}, \bibinfo {author} {\bibfnamefont {E.}~\bibnamefont
  {Ruiz~Arriola}}, \ and\ \bibinfo {author} {\bibfnamefont {I.}~\bibnamefont
  {Ruiz~Simo}},\ }\href {\doibase 10.1103/PhysRevC.98.024627} {\bibfield
  {journal} {\bibinfo  {journal} {Phys. Rev. C}\ }\textbf {\bibinfo {volume}
  {98}},\ \bibinfo {pages} {024627} (\bibinfo {year} {2018})}\BibitemShut
  {NoStop}%
\bibitem [{\citenamefont {Mosel}\ and\ \citenamefont
  {Gallmeister}(2019)}]{Mosel19}%
  \BibitemOpen
  \bibfield  {author} {\bibinfo {author} {\bibfnamefont {U.}~\bibnamefont
  {Mosel}}\ and\ \bibinfo {author} {\bibfnamefont {K.}~\bibnamefont
  {Gallmeister}},\ }\href {\doibase 10.1103/PhysRevC.99.064605} {\bibfield
  {journal} {\bibinfo  {journal} {Phys. Rev. C}\ }\textbf {\bibinfo {volume}
  {99}},\ \bibinfo {pages} {064605} (\bibinfo {year} {2019})}\BibitemShut
  {NoStop}%
\bibitem [{\citenamefont {Barbaro}\ \emph {et~al.}(2019)\citenamefont
  {Barbaro}, \citenamefont {Caballero}, \citenamefont {De~Pace}, \citenamefont
  {Donnelly}, \citenamefont {Gonz\'alez-Jim\'enez},\ and\ \citenamefont
  {Megias}}]{Barbaro19}%
  \BibitemOpen
  \bibfield  {author} {\bibinfo {author} {\bibfnamefont {M.~B.}\ \bibnamefont
  {Barbaro}}, \bibinfo {author} {\bibfnamefont {J.~A.}\ \bibnamefont
  {Caballero}}, \bibinfo {author} {\bibfnamefont {A.}~\bibnamefont {De~Pace}},
  \bibinfo {author} {\bibfnamefont {T.~W.}\ \bibnamefont {Donnelly}}, \bibinfo
  {author} {\bibfnamefont {R.}~\bibnamefont {Gonz\'alez-Jim\'enez}}, \ and\
  \bibinfo {author} {\bibfnamefont {G.~D.}\ \bibnamefont {Megias}},\ }\href
  {\doibase 10.1103/PhysRevC.99.042501} {\bibfield  {journal} {\bibinfo
  {journal} {Phys. Rev. C}\ }\textbf {\bibinfo {volume} {99}},\ \bibinfo
  {pages} {042501} (\bibinfo {year} {2019})}\BibitemShut {NoStop}%
\bibitem [{\citenamefont {Gonz\'alez-Jim\'enez}\ \emph
  {et~al.}(2019)\citenamefont {Gonz\'alez-Jim\'enez}, \citenamefont
  {Nikolakopoulos}, \citenamefont {Jachowicz},\ and\ \citenamefont
  {Ud\'{\i}as}}]{Gonzalez-Jimenez19}%
  \BibitemOpen
  \bibfield  {author} {\bibinfo {author} {\bibfnamefont {R.}~\bibnamefont
  {Gonz\'alez-Jim\'enez}}, \bibinfo {author} {\bibfnamefont {A.}~\bibnamefont
  {Nikolakopoulos}}, \bibinfo {author} {\bibfnamefont {N.}~\bibnamefont
  {Jachowicz}}, \ and\ \bibinfo {author} {\bibfnamefont {J.~M.}\ \bibnamefont
  {Ud\'{\i}as}},\ }\href {\doibase 10.1103/PhysRevC.100.045501} {\bibfield
  {journal} {\bibinfo  {journal} {Phys. Rev. C}\ }\textbf {\bibinfo {volume}
  {100}},\ \bibinfo {pages} {045501} (\bibinfo {year} {2019})}\BibitemShut
  {NoStop}%
\bibitem [{\citenamefont {Cooper}\ \emph {et~al.}(1993)\citenamefont {Cooper},
  \citenamefont {Hama}, \citenamefont {Clark},\ and\ \citenamefont
  {Mercer}}]{Cooper93}%
  \BibitemOpen
  \bibfield  {author} {\bibinfo {author} {\bibfnamefont {E.~D.}\ \bibnamefont
  {Cooper}}, \bibinfo {author} {\bibfnamefont {S.}~\bibnamefont {Hama}},
  \bibinfo {author} {\bibfnamefont {B.~C.}\ \bibnamefont {Clark}}, \ and\
  \bibinfo {author} {\bibfnamefont {R.~L.}\ \bibnamefont {Mercer}},\ }\href
  {\doibase 10.1103/PhysRevC.47.297} {\bibfield  {journal} {\bibinfo  {journal}
  {Phys. Rev. C}\ }\textbf {\bibinfo {volume} {47}},\ \bibinfo {pages} {297}
  (\bibinfo {year} {1993})}\BibitemShut {NoStop}%
\bibitem [{\citenamefont {Cooper}\ \emph {et~al.}(2009)\citenamefont {Cooper},
  \citenamefont {Hama},\ and\ \citenamefont {Clark}}]{Cooper09}%
  \BibitemOpen
  \bibfield  {author} {\bibinfo {author} {\bibfnamefont {E.~D.}\ \bibnamefont
  {Cooper}}, \bibinfo {author} {\bibfnamefont {S.}~\bibnamefont {Hama}}, \ and\
  \bibinfo {author} {\bibfnamefont {B.~C.}\ \bibnamefont {Clark}},\ }\href
  {\doibase 10.1103/PhysRevC.80.034605} {\bibfield  {journal} {\bibinfo
  {journal} {Phys. Rev. C}\ }\textbf {\bibinfo {volume} {80}},\ \bibinfo
  {pages} {034605} (\bibinfo {year} {2009})}\BibitemShut {NoStop}%
\bibitem [{\citenamefont {Maieron}\ \emph {et~al.}(2003)\citenamefont
  {Maieron}, \citenamefont {Mart\'{\i}nez}, \citenamefont {Caballero},\ and\
  \citenamefont {Ud\'{\i}as}}]{Maieron03}%
  \BibitemOpen
  \bibfield  {author} {\bibinfo {author} {\bibfnamefont {C.}~\bibnamefont
  {Maieron}}, \bibinfo {author} {\bibfnamefont {M.~C.}\ \bibnamefont
  {Mart\'{\i}nez}}, \bibinfo {author} {\bibfnamefont {J.~A.}\ \bibnamefont
  {Caballero}}, \ and\ \bibinfo {author} {\bibfnamefont {J.~M.}\ \bibnamefont
  {Ud\'{\i}as}},\ }\href {\doibase 10.1103/PhysRevC.68.048501} {\bibfield
  {journal} {\bibinfo  {journal} {Phys. Rev. C}\ }\textbf {\bibinfo {volume}
  {68}},\ \bibinfo {pages} {048501} (\bibinfo {year} {2003})}\BibitemShut
  {NoStop}%
\bibitem [{\citenamefont {Kim}\ and\ \citenamefont {Wright}(2003)}]{Kim03}%
  \BibitemOpen
  \bibfield  {author} {\bibinfo {author} {\bibfnamefont {K.~S.}\ \bibnamefont
  {Kim}}\ and\ \bibinfo {author} {\bibfnamefont {L.~E.}\ \bibnamefont
  {Wright}},\ }\href {\doibase 10.1103/PhysRevC.68.027601} {\bibfield
  {journal} {\bibinfo  {journal} {Phys. Rev. C}\ }\textbf {\bibinfo {volume}
  {68}},\ \bibinfo {pages} {027601} (\bibinfo {year} {2003})}\BibitemShut
  {NoStop}%
\bibitem [{\citenamefont {Caballero}\ \emph {et~al.}(2005)\citenamefont
  {Caballero}, \citenamefont {Amaro}, \citenamefont {Barbaro}, \citenamefont
  {Donnelly}, \citenamefont {Maieron},\ and\ \citenamefont
  {Udias}}]{Caballero05}%
  \BibitemOpen
  \bibfield  {author} {\bibinfo {author} {\bibfnamefont {J.~A.}\ \bibnamefont
  {Caballero}}, \bibinfo {author} {\bibfnamefont {J.~E.}\ \bibnamefont
  {Amaro}}, \bibinfo {author} {\bibfnamefont {M.~B.}\ \bibnamefont {Barbaro}},
  \bibinfo {author} {\bibfnamefont {T.~W.}\ \bibnamefont {Donnelly}}, \bibinfo
  {author} {\bibfnamefont {C.}~\bibnamefont {Maieron}}, \ and\ \bibinfo
  {author} {\bibfnamefont {J.~M.}\ \bibnamefont {Udias}},\ }\href {\doibase
  10.1103/PhysRevLett.95.252502} {\bibfield  {journal} {\bibinfo  {journal}
  {Phys. Rev. Lett.}\ }\textbf {\bibinfo {volume} {95}},\ \bibinfo {pages}
  {252502} (\bibinfo {year} {2005})}\BibitemShut {NoStop}%
\bibitem [{\citenamefont {Kim}\ and\ \citenamefont {Wright}(2007)}]{Kim07}%
  \BibitemOpen
  \bibfield  {author} {\bibinfo {author} {\bibfnamefont {K.~S.}\ \bibnamefont
  {Kim}}\ and\ \bibinfo {author} {\bibfnamefont {L.~E.}\ \bibnamefont
  {Wright}},\ }\href {\doibase 10.1103/PhysRevC.76.044613} {\bibfield
  {journal} {\bibinfo  {journal} {Phys. Rev. C}\ }\textbf {\bibinfo {volume}
  {76}},\ \bibinfo {pages} {044613} (\bibinfo {year} {2007})}\BibitemShut
  {NoStop}%
\bibitem [{\citenamefont {Butkevich}\ and\ \citenamefont
  {Kulagin}(2007)}]{Butkevich07}%
  \BibitemOpen
  \bibfield  {author} {\bibinfo {author} {\bibfnamefont {A.~V.}\ \bibnamefont
  {Butkevich}}\ and\ \bibinfo {author} {\bibfnamefont {S.~A.}\ \bibnamefont
  {Kulagin}},\ }\href {\doibase 10.1103/PhysRevC.76.045502} {\bibfield
  {journal} {\bibinfo  {journal} {Phys. Rev. C}\ }\textbf {\bibinfo {volume}
  {76}},\ \bibinfo {pages} {045502} (\bibinfo {year} {2007})}\BibitemShut
  {NoStop}%
\bibitem [{\citenamefont {Megias}\ \emph
  {et~al.}(2016{\natexlab{a}})\citenamefont {Megias}, \citenamefont {Amaro},
  \citenamefont {Barbaro}, \citenamefont {Caballero},\ and\ \citenamefont
  {Donnelly}}]{Megias16a}%
  \BibitemOpen
  \bibfield  {author} {\bibinfo {author} {\bibfnamefont {G.~D.}\ \bibnamefont
  {Megias}}, \bibinfo {author} {\bibfnamefont {J.~E.}\ \bibnamefont {Amaro}},
  \bibinfo {author} {\bibfnamefont {M.~B.}\ \bibnamefont {Barbaro}}, \bibinfo
  {author} {\bibfnamefont {J.~A.}\ \bibnamefont {Caballero}}, \ and\ \bibinfo
  {author} {\bibfnamefont {T.~W.}\ \bibnamefont {Donnelly}},\ }\href {\doibase
  10.1103/PhysRevD.94.013012} {\bibfield  {journal} {\bibinfo  {journal} {Phys.
  Rev. D}\ }\textbf {\bibinfo {volume} {94}},\ \bibinfo {pages} {013012}
  (\bibinfo {year} {2016}{\natexlab{a}})}\BibitemShut {NoStop}%
\bibitem [{\citenamefont {Pandey}\ \emph {et~al.}(2015)\citenamefont {Pandey},
  \citenamefont {Jachowicz}, \citenamefont {Van~Cuyck}, \citenamefont
  {Ryckebusch},\ and\ \citenamefont {Martini}}]{Pandey15}%
  \BibitemOpen
  \bibfield  {author} {\bibinfo {author} {\bibfnamefont {V.}~\bibnamefont
  {Pandey}}, \bibinfo {author} {\bibfnamefont {N.}~\bibnamefont {Jachowicz}},
  \bibinfo {author} {\bibfnamefont {T.}~\bibnamefont {Van~Cuyck}}, \bibinfo
  {author} {\bibfnamefont {J.}~\bibnamefont {Ryckebusch}}, \ and\ \bibinfo
  {author} {\bibfnamefont {M.}~\bibnamefont {Martini}},\ }\href {\doibase
  10.1103/PhysRevC.92.024606} {\bibfield  {journal} {\bibinfo  {journal} {Phys.
  Rev. C}\ }\textbf {\bibinfo {volume} {92}},\ \bibinfo {pages} {024606}
  (\bibinfo {year} {2015})}\BibitemShut {NoStop}%
\bibitem [{\citenamefont {Pandey}\ \emph {et~al.}(2016)\citenamefont {Pandey},
  \citenamefont {Jachowicz}, \citenamefont {Martini}, \citenamefont
  {Gonz\'alez-Jim\'enez}, \citenamefont {Ryckebusch}, \citenamefont
  {Van~Cuyck},\ and\ \citenamefont {Van~Dessel}}]{Pandey16}%
  \BibitemOpen
  \bibfield  {author} {\bibinfo {author} {\bibfnamefont {V.}~\bibnamefont
  {Pandey}}, \bibinfo {author} {\bibfnamefont {N.}~\bibnamefont {Jachowicz}},
  \bibinfo {author} {\bibfnamefont {M.}~\bibnamefont {Martini}}, \bibinfo
  {author} {\bibfnamefont {R.}~\bibnamefont {Gonz\'alez-Jim\'enez}}, \bibinfo
  {author} {\bibfnamefont {J.}~\bibnamefont {Ryckebusch}}, \bibinfo {author}
  {\bibfnamefont {T.}~\bibnamefont {Van~Cuyck}}, \ and\ \bibinfo {author}
  {\bibfnamefont {N.}~\bibnamefont {Van~Dessel}},\ }\href {\doibase
  10.1103/PhysRevC.94.054609} {\bibfield  {journal} {\bibinfo  {journal} {Phys.
  Rev. C}\ }\textbf {\bibinfo {volume} {94}},\ \bibinfo {pages} {054609}
  (\bibinfo {year} {2016})}\BibitemShut {NoStop}%
\bibitem [{\citenamefont {Waroquier}\ \emph {et~al.}(1987)\citenamefont
  {Waroquier}, \citenamefont {Ryckebusch}, \citenamefont {Moreau},
  \citenamefont {Heyde}, \citenamefont {Blasi}, \citenamefont {van~der Werf},\
  and\ \citenamefont {Wenes}}]{Waroquier:1986mj}%
  \BibitemOpen
  \bibfield  {author} {\bibinfo {author} {\bibfnamefont {M.}~\bibnamefont
  {Waroquier}}, \bibinfo {author} {\bibfnamefont {J.}~\bibnamefont
  {Ryckebusch}}, \bibinfo {author} {\bibfnamefont {J.}~\bibnamefont {Moreau}},
  \bibinfo {author} {\bibfnamefont {K.}~\bibnamefont {Heyde}}, \bibinfo
  {author} {\bibfnamefont {N.}~\bibnamefont {Blasi}}, \bibinfo {author}
  {\bibfnamefont {S.~Y.}\ \bibnamefont {van~der Werf}}, \ and\ \bibinfo
  {author} {\bibfnamefont {G.}~\bibnamefont {Wenes}},\ }\href {\doibase
  10.1016/0370-1573(87)90066-4} {\bibfield  {journal} {\bibinfo  {journal}
  {Phys. Rept.}\ }\textbf {\bibinfo {volume} {148}},\ \bibinfo {pages} {249}
  (\bibinfo {year} {1987})}\BibitemShut {NoStop}%
\bibitem [{\citenamefont {Walecka}(2004)}]{walecka04}%
  \BibitemOpen
  \bibfield  {author} {\bibinfo {author} {\bibfnamefont {J.}~\bibnamefont
  {Walecka}},\ }\href {https://books.google.be/books?id=sBABwQEACAAJ} {\emph
  {\bibinfo {title} {Theoretical Nuclear and Subnuclear Physics}}}\ (\bibinfo
  {publisher} {World Scientfic, Imperial College Press},\ \bibinfo {year}
  {2004})\BibitemShut {NoStop}%
\bibitem [{\citenamefont {Fetter}\ and\ \citenamefont
  {Walecka}(1971)}]{fetterwalecka1971}%
  \BibitemOpen
  \bibfield  {author} {\bibinfo {author} {\bibfnamefont {A.}~\bibnamefont
  {Fetter}}\ and\ \bibinfo {author} {\bibfnamefont {J.}~\bibnamefont
  {Walecka}},\ }\href {https://books.google.be/books?id=0wekf1s83b0C} {\emph
  {\bibinfo {title} {Quantum Theory of Many{--}Particle Systems}}}\ (\bibinfo
  {publisher} {Dover Publications, inc.},\ \bibinfo {year} {1971})\BibitemShut
  {NoStop}%
\bibitem [{\citenamefont {Jeschonnek}\ and\ \citenamefont
  {Donnelly}(1998)}]{Jeschonnek}%
  \BibitemOpen
  \bibfield  {author} {\bibinfo {author} {\bibfnamefont {S.}~\bibnamefont
  {Jeschonnek}}\ and\ \bibinfo {author} {\bibfnamefont {T.~W.}\ \bibnamefont
  {Donnelly}},\ }\href {\doibase 10.1103/PhysRevC.57.2438} {\bibfield
  {journal} {\bibinfo  {journal} {Phys. Rev. C}\ }\textbf {\bibinfo {volume}
  {57}},\ \bibinfo {pages} {2438} (\bibinfo {year} {1998})}\BibitemShut
  {NoStop}%
\bibitem [{\citenamefont {Jachowicz}\ \emph {et~al.}(2019)\citenamefont
  {Jachowicz}, \citenamefont {Dessel},\ and\ \citenamefont
  {Nikolakopoulos}}]{Jachowicz19}%
  \BibitemOpen
  \bibfield  {author} {\bibinfo {author} {\bibfnamefont {N.}~\bibnamefont
  {Jachowicz}}, \bibinfo {author} {\bibfnamefont {N.~V.}\ \bibnamefont
  {Dessel}}, \ and\ \bibinfo {author} {\bibfnamefont {A.}~\bibnamefont
  {Nikolakopoulos}},\ }\href {\doibase 10.1088/1361-6471/ab25d4} {\bibfield
  {journal} {\bibinfo  {journal} {Journal of Physics G: Nuclear and Particle
  Physics}\ }\textbf {\bibinfo {volume} {46}},\ \bibinfo {pages} {084003}
  (\bibinfo {year} {2019})}\BibitemShut {NoStop}%
\bibitem [{\citenamefont {Donnelly}\ and\ \citenamefont
  {Sick}(1999{\natexlab{a}})}]{Donnelly99a}%
  \BibitemOpen
  \bibfield  {author} {\bibinfo {author} {\bibfnamefont {T.~W.}\ \bibnamefont
  {Donnelly}}\ and\ \bibinfo {author} {\bibfnamefont {I.}~\bibnamefont
  {Sick}},\ }\href {\doibase 10.1103/PhysRevLett.82.3212} {\bibfield  {journal}
  {\bibinfo  {journal} {Phys. Rev. Lett.}\ }\textbf {\bibinfo {volume} {82}},\
  \bibinfo {pages} {3212} (\bibinfo {year} {1999}{\natexlab{a}})}\BibitemShut
  {NoStop}%
\bibitem [{\citenamefont {Donnelly}\ and\ \citenamefont
  {Sick}(1999{\natexlab{b}})}]{Donnelly99b}%
  \BibitemOpen
  \bibfield  {author} {\bibinfo {author} {\bibfnamefont {T.~W.}\ \bibnamefont
  {Donnelly}}\ and\ \bibinfo {author} {\bibfnamefont {I.}~\bibnamefont
  {Sick}},\ }\href {\doibase 10.1103/PhysRevC.60.065502} {\bibfield  {journal}
  {\bibinfo  {journal} {Phys. Rev. C}\ }\textbf {\bibinfo {volume} {60}},\
  \bibinfo {pages} {065502} (\bibinfo {year} {1999}{\natexlab{b}})}\BibitemShut
  {NoStop}%
\bibitem [{\citenamefont {Barbaro}\ \emph {et~al.}(2004)\citenamefont
  {Barbaro}, \citenamefont {Caballero}, \citenamefont {Donnelly},\ and\
  \citenamefont {Maieron}}]{Barbaro04}%
  \BibitemOpen
  \bibfield  {author} {\bibinfo {author} {\bibfnamefont {M.~B.}\ \bibnamefont
  {Barbaro}}, \bibinfo {author} {\bibfnamefont {J.~A.}\ \bibnamefont
  {Caballero}}, \bibinfo {author} {\bibfnamefont {T.~W.}\ \bibnamefont
  {Donnelly}}, \ and\ \bibinfo {author} {\bibfnamefont {C.}~\bibnamefont
  {Maieron}},\ }\href {\doibase 10.1103/PhysRevC.69.035502} {\bibfield
  {journal} {\bibinfo  {journal} {Phys. Rev. C}\ }\textbf {\bibinfo {volume}
  {69}},\ \bibinfo {pages} {035502} (\bibinfo {year} {2004})}\BibitemShut
  {NoStop}%
\bibitem [{\citenamefont {Maieron}\ \emph {et~al.}(2002)\citenamefont
  {Maieron}, \citenamefont {Donnelly},\ and\ \citenamefont {Sick}}]{Maieron02}%
  \BibitemOpen
  \bibfield  {author} {\bibinfo {author} {\bibfnamefont {C.}~\bibnamefont
  {Maieron}}, \bibinfo {author} {\bibfnamefont {T.~W.}\ \bibnamefont
  {Donnelly}}, \ and\ \bibinfo {author} {\bibfnamefont {I.}~\bibnamefont
  {Sick}},\ }\href {\doibase 10.1103/PhysRevC.65.025502} {\bibfield  {journal}
  {\bibinfo  {journal} {Phys. Rev. C}\ }\textbf {\bibinfo {volume} {65}},\
  \bibinfo {pages} {025502} (\bibinfo {year} {2002})}\BibitemShut {NoStop}%
\bibitem [{\citenamefont {Maieron}\ \emph {et~al.}(2009)\citenamefont
  {Maieron}, \citenamefont {Amaro}, \citenamefont {Barbaro}, \citenamefont
  {Caballero}, \citenamefont {Donnelly},\ and\ \citenamefont
  {Williamson}}]{Maieron09}%
  \BibitemOpen
  \bibfield  {author} {\bibinfo {author} {\bibfnamefont {C.}~\bibnamefont
  {Maieron}}, \bibinfo {author} {\bibfnamefont {J.~E.}\ \bibnamefont {Amaro}},
  \bibinfo {author} {\bibfnamefont {M.~B.}\ \bibnamefont {Barbaro}}, \bibinfo
  {author} {\bibfnamefont {J.~A.}\ \bibnamefont {Caballero}}, \bibinfo {author}
  {\bibfnamefont {T.~W.}\ \bibnamefont {Donnelly}}, \ and\ \bibinfo {author}
  {\bibfnamefont {C.~F.}\ \bibnamefont {Williamson}},\ }\href {\doibase
  10.1103/PhysRevC.80.035504} {\bibfield  {journal} {\bibinfo  {journal} {Phys.
  Rev. C}\ }\textbf {\bibinfo {volume} {80}},\ \bibinfo {pages} {035504}
  (\bibinfo {year} {2009})}\BibitemShut {NoStop}%
\bibitem [{\citenamefont {Megias}\ \emph {et~al.}(2015)\citenamefont {Megias},
  \citenamefont {Donnelly}, \citenamefont {Moreno}, \citenamefont {Williamson},
  \citenamefont {Caballero}, \citenamefont {Gonz\'alez-Jim\'enez},
  \citenamefont {{De Pace}}, \citenamefont {Barbaro}, \citenamefont {Alberico},
  \citenamefont {Nardi},\ and\ \citenamefont {Amaro}}]{Megias15}%
  \BibitemOpen
  \bibfield  {author} {\bibinfo {author} {\bibfnamefont {G.}~\bibnamefont
  {Megias}}, \bibinfo {author} {\bibfnamefont {T.}~\bibnamefont {Donnelly}},
  \bibinfo {author} {\bibfnamefont {O.}~\bibnamefont {Moreno}}, \bibinfo
  {author} {\bibfnamefont {C.}~\bibnamefont {Williamson}}, \bibinfo {author}
  {\bibfnamefont {J.}~\bibnamefont {Caballero}}, \bibinfo {author}
  {\bibfnamefont {R.}~\bibnamefont {Gonz\'alez-Jim\'enez}}, \bibinfo {author}
  {\bibfnamefont {A.}~\bibnamefont {{De Pace}}}, \bibinfo {author}
  {\bibfnamefont {M.}~\bibnamefont {Barbaro}}, \bibinfo {author} {\bibfnamefont
  {W.}~\bibnamefont {Alberico}}, \bibinfo {author} {\bibfnamefont
  {M.}~\bibnamefont {Nardi}}, \ and\ \bibinfo {author} {\bibfnamefont
  {J.}~\bibnamefont {Amaro}},\ }\href {\doibase 10.1103/PhysRevD.91.073004}
  {\bibfield  {journal} {\bibinfo  {journal} {Phys. Rev. D}\ }\textbf {\bibinfo
  {volume} {91}},\ \bibinfo {pages} {073004} (\bibinfo {year}
  {2015})}\BibitemShut {NoStop}%
\bibitem [{\citenamefont {Megias}\ \emph
  {et~al.}(2016{\natexlab{b}})\citenamefont {Megias}, \citenamefont {Amaro},
  \citenamefont {Barbaro}, \citenamefont {Caballero}, \citenamefont
  {Donnelly},\ and\ \citenamefont {Simo}}]{Megias16b}%
  \BibitemOpen
  \bibfield  {author} {\bibinfo {author} {\bibfnamefont {G.~D.}\ \bibnamefont
  {Megias}}, \bibinfo {author} {\bibfnamefont {J.~E.}\ \bibnamefont {Amaro}},
  \bibinfo {author} {\bibfnamefont {M.~B.}\ \bibnamefont {Barbaro}}, \bibinfo
  {author} {\bibfnamefont {J.~A.}\ \bibnamefont {Caballero}}, \bibinfo {author}
  {\bibfnamefont {T.~W.}\ \bibnamefont {Donnelly}}, \ and\ \bibinfo {author}
  {\bibfnamefont {I.~R.}\ \bibnamefont {Simo}},\ }\href {\doibase
  10.1103/PhysRevD.94.093004} {\bibfield  {journal} {\bibinfo  {journal} {Phys.
  Rev. D}\ }\textbf {\bibinfo {volume} {94}},\ \bibinfo {pages} {093004}
  (\bibinfo {year} {2016}{\natexlab{b}})}\BibitemShut {NoStop}%
\bibitem [{\citenamefont {Megias}\ \emph {et~al.}(2014)\citenamefont {Megias},
  \citenamefont {Ivanov}, \citenamefont {Gonz\'alez-Jim\'enez}, \citenamefont
  {Barbaro}, \citenamefont {Caballero}, \citenamefont {Donnelly},\ and\
  \citenamefont {Ud\'{\i}as}}]{Megias14}%
  \BibitemOpen
  \bibfield  {author} {\bibinfo {author} {\bibfnamefont {G.~D.}\ \bibnamefont
  {Megias}}, \bibinfo {author} {\bibfnamefont {M.~V.}\ \bibnamefont {Ivanov}},
  \bibinfo {author} {\bibfnamefont {R.}~\bibnamefont {Gonz\'alez-Jim\'enez}},
  \bibinfo {author} {\bibfnamefont {M.~B.}\ \bibnamefont {Barbaro}}, \bibinfo
  {author} {\bibfnamefont {J.~A.}\ \bibnamefont {Caballero}}, \bibinfo {author}
  {\bibfnamefont {T.~W.}\ \bibnamefont {Donnelly}}, \ and\ \bibinfo {author}
  {\bibfnamefont {J.~M.}\ \bibnamefont {Ud\'{\i}as}},\ }\href {\doibase
  10.1103/PhysRevD.89.093002} {\bibfield  {journal} {\bibinfo  {journal} {Phys.
  Rev. D}\ }\textbf {\bibinfo {volume} {89}},\ \bibinfo {pages} {093002}
  (\bibinfo {year} {2014})}\BibitemShut {NoStop}%
\bibitem [{\citenamefont {Acciarri}\ \emph {et~al.}(2016)\citenamefont
  {Acciarri} \emph {et~al.}}]{DUNE16}%
  \BibitemOpen
  \bibfield  {author} {\bibinfo {author} {\bibfnamefont {R.}~\bibnamefont
  {Acciarri}} \emph {et~al.} (\bibinfo {collaboration} {DUNE}),\ }\href@noop {}
  {\  (\bibinfo {year} {2016})},\ \Eprint {http://arxiv.org/abs/1601.05471}
  {arXiv:1601.05471 [physics.ins-det]} \BibitemShut {NoStop}%
\bibitem [{\citenamefont {Dai}\ \emph {et~al.}(2019)\citenamefont {Dai} \emph
  {et~al.}}]{JLab_Ar40}%
  \BibitemOpen
  \bibfield  {author} {\bibinfo {author} {\bibfnamefont {H.}~\bibnamefont
  {Dai}} \emph {et~al.} (\bibinfo {collaboration} {The Jefferson Lab Hall A
  Collaboration}),\ }\href {\doibase 10.1103/PhysRevC.99.054608} {\bibfield
  {journal} {\bibinfo  {journal} {Phys. Rev. C}\ }\textbf {\bibinfo {volume}
  {99}},\ \bibinfo {pages} {054608} (\bibinfo {year} {2019})}\BibitemShut
  {NoStop}%
\bibitem [{\citenamefont {Dai}\ \emph {et~al.}(2018)\citenamefont {Dai} \emph
  {et~al.}}]{JLab_Ti48}%
  \BibitemOpen
  \bibfield  {author} {\bibinfo {author} {\bibfnamefont {H.}~\bibnamefont
  {Dai}} \emph {et~al.} (\bibinfo {collaboration} {Jefferson Lab Hall A
  Collaboration}),\ }\href {\doibase 10.1103/PhysRevC.98.014617} {\bibfield
  {journal} {\bibinfo  {journal} {Phys. Rev. C}\ }\textbf {\bibinfo {volume}
  {98}},\ \bibinfo {pages} {014617} (\bibinfo {year} {2018})}\BibitemShut
  {NoStop}%
\bibitem [{\citenamefont {Gonz\'alez-Jim\'enez}\ \emph
  {et~al.}(2017)\citenamefont {Gonz\'alez-Jim\'enez}, \citenamefont
  {Jachowicz}, \citenamefont {Niewczas}, \citenamefont {Nys}, \citenamefont
  {Pandey}, \citenamefont {Van~Cuyck},\ and\ \citenamefont
  {Van~Dessel}}]{Gonzalez-Jimenez17}%
  \BibitemOpen
  \bibfield  {author} {\bibinfo {author} {\bibfnamefont {R.}~\bibnamefont
  {Gonz\'alez-Jim\'enez}}, \bibinfo {author} {\bibfnamefont {N.}~\bibnamefont
  {Jachowicz}}, \bibinfo {author} {\bibfnamefont {K.}~\bibnamefont {Niewczas}},
  \bibinfo {author} {\bibfnamefont {J.}~\bibnamefont {Nys}}, \bibinfo {author}
  {\bibfnamefont {V.}~\bibnamefont {Pandey}}, \bibinfo {author} {\bibfnamefont
  {T.}~\bibnamefont {Van~Cuyck}}, \ and\ \bibinfo {author} {\bibfnamefont
  {N.}~\bibnamefont {Van~Dessel}},\ }\href {\doibase
  10.1103/PhysRevD.95.113007} {\bibfield  {journal} {\bibinfo  {journal} {Phys.
  Rev. D}\ }\textbf {\bibinfo {volume} {95}},\ \bibinfo {pages} {113007}
  (\bibinfo {year} {2017})}\BibitemShut {NoStop}%
\bibitem [{\citenamefont {Gonz\'alez-Jim\'enez}\ \emph
  {et~al.}(2018)\citenamefont {Gonz\'alez-Jim\'enez}, \citenamefont
  {Niewczas},\ and\ \citenamefont {Jachowicz}}]{Gonzalez-Jimenez18}%
  \BibitemOpen
  \bibfield  {author} {\bibinfo {author} {\bibfnamefont {R.}~\bibnamefont
  {Gonz\'alez-Jim\'enez}}, \bibinfo {author} {\bibfnamefont {K.}~\bibnamefont
  {Niewczas}}, \ and\ \bibinfo {author} {\bibfnamefont {N.}~\bibnamefont
  {Jachowicz}},\ }\href {\doibase 10.1103/PhysRevD.97.013004} {\bibfield
  {journal} {\bibinfo  {journal} {Phys. Rev. D}\ }\textbf {\bibinfo {volume}
  {97}},\ \bibinfo {pages} {013004} (\bibinfo {year} {2018})}\BibitemShut
  {NoStop}%
\bibitem [{\citenamefont {Nikolakopoulos}\ \emph {et~al.}(2018)\citenamefont
  {Nikolakopoulos}, \citenamefont {Gonz\'alez-Jim\'enez}, \citenamefont
  {Niewczas}, \citenamefont {Sobczyk},\ and\ \citenamefont
  {Jachowicz}}]{Nikolakopoulos18a}%
  \BibitemOpen
  \bibfield  {author} {\bibinfo {author} {\bibfnamefont {A.}~\bibnamefont
  {Nikolakopoulos}}, \bibinfo {author} {\bibfnamefont {R.}~\bibnamefont
  {Gonz\'alez-Jim\'enez}}, \bibinfo {author} {\bibfnamefont {K.}~\bibnamefont
  {Niewczas}}, \bibinfo {author} {\bibfnamefont {J.}~\bibnamefont {Sobczyk}}, \
  and\ \bibinfo {author} {\bibfnamefont {N.}~\bibnamefont {Jachowicz}},\ }\href
  {\doibase 10.1103/PhysRevD.97.093008} {\bibfield  {journal} {\bibinfo
  {journal} {Phys. Rev. D}\ }\textbf {\bibinfo {volume} {97}},\ \bibinfo
  {pages} {093008} (\bibinfo {year} {2018})}\BibitemShut {NoStop}%
\end{thebibliography}%
}

\end{document}